\documentclass[aps,prb,amsmath,amssymb,twocolumn,superscriptaddress,showpacs]{revtex4-2}

\usepackage[english]{babel}
\usepackage[letterpaper,top=2cm,bottom=2cm,left=3cm,right=3cm,marginparwidth=1.75cm]{geometry}

\usepackage{amsmath}
\usepackage{graphicx}
\usepackage[colorlinks=true, allcolors=blue]{hyperref}
\usepackage{physics}

\begin{document}
%\title{Nonreciprocial dynamics in Raman and Brillouin scattering}
%\title{Raman scattering from a quantum field perspective}
\title{Entanglement and Optical Nonreciprocity in spontaneous Raman Scattering}

\author{Frank Schlawin}
\affiliation{Max Planck Institute for the Structure and Dynamics of Matter, Luruper Chaussee 149, 22761 Hamburg, Germany}
\affiliation{University of Hamburg, Luruper Chaussee 149, Hamburg, Germany}
\affiliation{The Hamburg Centre for Ultrafast Imaging, Luruper Chaussee 149, Hamburg D-22761, Germany}

\begin{abstract}

Inelastic light scattering is a central tool for sample characterization and label-free imaging across the physical and life sciences. Recent work has suggested that scattered light can also exhibit nonclassical correlations. Here, we develop a microscopic theory of Raman scattering that connects naturally to established descriptions of entangled-photon generation in three- and four-wave mixing, while capturing essential differences arising from the resonant, dissipative character of Raman scattering. Using a cumulant-expansion approach, we analyze the entanglement structure of the scattered sidebands and identify signatures that distinguish Raman-mediated correlations from those generated in conventional off-resonant nonlinear wave mixing. In particular, we show that Raman scattering induces chiral couplings between Stokes and anti-Stokes sidebands, leading to nonreciprocal amplification of coherent seed fields. These results establish a theoretical framework for Raman-based quantum photonic protocols and suggest routes toward quantum-enhanced Raman spectroscopy and imaging.

\end{abstract}
\maketitle

\section{Introduction}

Raman scattering is a ubiquitous tool in research ranging from life sciences~\cite{Cheng2015} to condensed matter physics~\cite{Ma2021}. 
It is an inelastic light scattering process that produces two sidebands: the Stokes field, in which an incident photon creates a phonon in the sample and loses energy, and the anti-Stokes field, in which a phonon is annihilated and the scattered photon gains the energy of the annihilated phonon. This familiar semiclassical picture is, however, incomplete: As first pointed out by Klyshko~\cite{Klyshko1977}, two scattering events may first generate and then destroy a single phonon in the sample, and thereby generate a Stokes and an anti-Stokes photon in a quantum correlated pair, which can be inferred from nonclassical correlations between the two fields. 
%, implying nonclassical correlations between the two emitted fields. 
This prediction has since been confirmed experimentally in a number of experiments~\cite{Kaperczyk2016, Saraiva2017, Anderson2018, Velez2019, Velez2020, Shinbrough2020, Junior2019, Freitas2024, Aguiar2024}. Although the Stokes and anti-Stokes beams individually exhibit thermal statistics, their joint measurements reveal strong pair correlations~\cite{Velez2019, Velez2020} and demonstrate the collective character of Raman excitations~\cite{Vento2023}. These results have raised interest in Raman-based quantum sensing, spectroscopy, and imaging, as well as in correlation-enhanced approaches more generally~\cite{Yang2020, DorfmanPNAS, Tollerud2019,Li:22, Li2024, Li:24, Zhu:25}.

Theoretical work inspired by these experiments has explored the quantum properties of the Stokes and anti-Stokes fields~\cite{Parra-Murillo2016, Guimaraes2020, Zhang2020, Thapliyal2021, Sier2025, Correa_2025}. Most microscopic treatments describe the vibrational and optical degrees of freedom explicitly, i.e. within a Hamiltonian framework, where all relevant degrees of freedom are evolved explicitly. 
This enables a detailed account of phonon heating and its interplay with correlations. However, such explicit approaches are naturally limited to few-mode settings and do not readily incorporate propagation and phase matching effects, or the spatiotemporal structure of the emitted fields in extended samples. Effective descriptions of Raman scattering in terms of nonlinear susceptibilities have also been developed since the 1980's~\cite{Raymer1980, Raymer1985, Glerean2019}, but these do not directly provide a microscopic dynamical theory of the emitted photonic states.

In this work, we introduce an alternative formulation of quantum Raman scattering based on a cumulant expansion of the light–matter interaction Hamiltonian~\cite{schlawin2025theoryquantumenhancedstimulatedraman, JCP2025_MDspec}. This construction yields an effective dynamical map for the photonic sidebands and provides a unified description that can incorporate finite bandwidths, phase matching in macroscopic samples, spatiotemporal quantum correlations, and realistic molecular response functions. 
We employ this map to analyze the dynamics and entanglement structure of spontaneous and weakly seeded Raman scattering. 
A key result is that resonant Raman scattering~\footnote{The term 'resonant' Raman scattering is often used to describe experiments where the optical transitions are resonant. Here, 'resonance' refers to the two-photon Raman resonance.} is intrinsically dissipative and nonreciprocal. More precisely, we show that its dynamics can be represented by a cascaded master equation~\cite{Gardiner1993, Carmichael1993, Stannigel_2012}, thereby placing spontaneous Raman scattering in the broader context of chiral open quantum systems~\cite{Lodahl2017, chiral-review}. 
We emphasize that this insight does not depend on chiral samples, and it is therefore distinct from manifestly chiral effects in Raman scattering, such as Raman optical activity~\cite{ROA-review}. 

The perspective emerging from our formulation has several implications. First, it shows that resonant Raman scattering does not generate quadrature squeezing below the vacuum limit, despite the presence of nontrivial photon-pair correlations, in contrast to nonresonant three- and four-wave mixing processes. 
Second, the correlations are reduced by the presence of spontaneously scattered individual photons, but may be retrieved by photon counting. 
Third, the dependence of the Raman response on phonon occupation leads to optical nonreciprocity. We define a set of collective modes which couple unilaterally, i.e. one mode stimulates emission into the other, but not the other way around. 
This suggests possible applications in quantum information processing and nonreciprocal amplification~\cite{Lodahl2017, Lin2019}. Finally, the formalism provides a basis for quantum metrology in Raman imaging~\cite{sorelli2025ultimateresolutionlimitscoherent} and for quantum-enhanced Raman spectroscopy. Recent experiments~\cite{deAndrade:20, Casacio2021, Li:22, Xu:22} and theory~\cite{Dorfman2014, Dorfman2021, Zhang2022, Fan2024, Pinto2026} have demonstrated the potential of stimulating Raman processes with quantum light; however, this has not yet been connected to the generation of correlated photonic states within the scattering process itself. 
Our work establishes this connection and thereby enables the investigation of stimulation and read-out of quantum correlations in Raman scattering in a simple, fully quantum mechanical theory. 
%Our work establishes this connection and thereby advances a dynamical quantum theory of Raman scattering in extended media.

%The paper is organized as follows. In Sec.~\ref{sec:master-equation}, we derive the cascaded master equation governing spontaneous Raman scattering. In Sec.~\ref{sec:simulations}, we present simulations of the generated states of light, analyze the quantum correlations and the nonreciprocal nature of the coupling between Stokes and anti-Stokes fields, before concluding in Sec.~\ref{sec:conclusions}.

\section{Master equation}
\label{sec:master-equation}

%We consider the setup sketched in Fig.~\ref{fig:sketch}(a). A strong, coherent pump pulse, which we treat classically, induces Stokes-anti-Stokes-scattering in an ensemble of molecules. Measurements are carried out on the output light fields which we treat fully quantum mechanically. In this paper, we consider only one relevant phonon mode, but our theory may be straightforwardly extended to include any number of vibrations, vibrational relaxation, or nonlinearities. 

We consider the setup sketched in Fig.~\ref{fig:sketch}(a). A strong coherent pump pulse, treated classically, induces Stokes and anti-Stokes scattering in an ensemble of molecules. The output light fields composed of the two Raman sidebands are measured and are treated fully quantum mechanically. In this paper, we focus on a single relevant phonon mode; however, the theory can be extended straightforwardly to include multiple vibrational modes, vibrational relaxation, or phonon nonlinearities.

\subsection{Light-matter interaction and cumulant expansion}

%We consider a molecule at position $\vec{r}$ interacting with the strong pump pulse. This pulse is sufficiently detuned form any electronic transition such that its main response (apart from elastic scattering) is given by inelastic scattering and the concomitant creation or annihilation of phonons in the molecule.  
%Within the sample region, the interaction Hamiltonian then reads
%We write the interaction Hamiltonian for this process as
We first consider a molecule~\footnote{A similar derivation and a similar effective Hamiltonian can also be invoked for Raman spectroscopy of materials. } at position $\vec{r}$ interacting with the strong pump pulse. The pulse is assumed to be sufficiently detuned from any electronic transition such that, apart from elastic scattering, the dominant response is inelastic scattering accompanied by the creation or annihilation of phonons in the molecule. The interaction Hamiltonian for this process can be written as
\begin{widetext}
\begin{align}
    \hat{H}_I (\vec{r}, t) &= \alpha_{\mathrm{R}} E^\ast_{\mathrm{pu}} (\vec{r}, t) \left( \hat{b}^\dagger (t) \hat{E}^\dagger_{\mathrm{St}} (\vec{r}, t) + \hat{b} (t) \hat{E}^\dagger_{\mathrm{aS}} (\vec{r}, t) \right) + h.c., \label{eq:H_I}
\end{align}
\end{widetext}
where $\hat{b}$ ($\hat{b}^\dagger$) are the phonon annihilation (creation) operators, respectively. $E_{\mathrm{pu}}$ is the classical pump field amplitude, and $\hat{E}_{\mathrm{St / aS}}$ are the Stokes (anti-Stokes) field operators (see Appendix~\ref{sec:field-operator}), respectively. The parameter $\alpha_{\mathrm{R}}$ denotes the Raman polarizability of the molecule.  
The total interaction Hamiltonian is then given by 
$\hat{H}_I (t) = \int d^3 r \hat{H}_I (\vec{r}, t)$. 
Treating the pump field classically corresponds to the undepleted-pump approximation familiar from parametric down-conversion. 
%, where it is well established even in regimes involving large photon numbers. In the case of parametric down-conversion, 
There, this approximation has been investigated experimentally and numerically~\cite{Florez:20}, and has been shown to remain valid up to macroscopic numbers of down-converted photons.
%Treating the pump fields classically is known as the undepleted pump approximation in parametric downconversion, where it is well established, even in the description of large photon numbers. In the case of parametric downconversion, this approximation was investigated experimentally and numerically~\cite{Florez:20}, and shown to hold up to macroscopic numbers of downconverted photons. 

Initially, at $t = -\infty$, the molecules are in a thermal state $\hat{\varrho}_{\mathrm{mol}}$ and the light field is prepared in an input state $\hat{\varrho}_{\mathrm{in}}$, which we mostly take to be the vacuum, but seed pulses or quantum seed states of light could be incorporated in the formalism straightforwardly. The initial joint density matrix therefore reads $\hat{\varrho}_{\mathrm{in}} \otimes \hat{\varrho}_{\mathrm{mol}}$. 
We are interested in the quantum state of the light long after its interaction with the molecules, i.e. in the asymptotic limit $t \rightarrow \infty$. 
The photonic density matrix after interaction with the sample is obtained by tracing out the molecular degrees of freedom,
\begin{align} \label{eq:rho_out-def}
    \hat{\varrho}_{\mathrm{out}} &= \text{tr}_{\mathrm{mol}} \left\{ \mathcal{S} \hat{\varrho}_{\mathrm{in}} \otimes \hat{\varrho}_{\mathrm{mol}} \right\},
\end{align}
where the scattering matrix $\mathcal{S}$ is given by the Dyson series, 
\begin{align}
    \mathcal{S} &= \mathcal{T} \exp \left[ - \frac{i}{\hbar} \int_{-\infty}^\infty \!\!\! d\tau \; \hat{H}_{I,-} (\tau) \right].
\end{align}
Here, the '-' subscript denotes the commutator superoperator, i.e. $H_- X = H X - X H$, and $\mathcal{T}$ is the time-ordering operator. 
Eq.~(\ref{eq:rho_out-def}) thus defines a quantum map, 
\begin{align}
\hat{\varrho}_{\mathrm{out}} &= \Phi [ \hat{\varrho}_{\mathrm{in}}], 
\end{align}
which connects the initial photonic density matrix $\hat{\varrho}_{\mathrm{in}}$ with the final one, $\hat{\varrho}_{\mathrm{out}}$. 
We next carry out a cumulant expansion of this quantum map, as in Refs.~\cite{schlawin2025theoryquantumenhancedstimulatedraman, JCP2025_MDspec}, and write it in an exponential form,
\begin{align}
    \Phi &\approx \exp \left[ \sum_j \mathcal{K}_j \right].
\end{align}
The cumulants $\mathcal{K}_j$ are obtained from time-ordered correlation functions of the form tr$_{\mathrm{mol}} \{ \mathcal{T} \hat{H}_{I,-} (\vec{r}_1, \tau_1) \ldots \hat{H}_{I,-} (\vec{r}_n, \tau_n) \}$. After tracing out only the molecular subsystem, these expressions become superoperators acting on the photonic degrees of freedom. 
Once the cumulants have been derived, the quantum map, or equivalently $\hat{\varrho}_{\mathrm{out}}$, can be obtained by propagating the auxiliary evolution equation
\begin{align} \label{eq:auxilliary}
    \partial_x \hat{\varrho} &= \sum_j \mathcal{K}_j \hat{\varrho}
\end{align}
from $x = 0$ to $1$, with initial condition $\hat{\varrho} (0) = \hat{\varrho}_{\mathrm{in}}$. 
%, where only the molecular degrees of freedom are traced out. 

\subsection{Full master equation}

\begin{figure}[t]
    \centering
    \includegraphics[width=\linewidth]{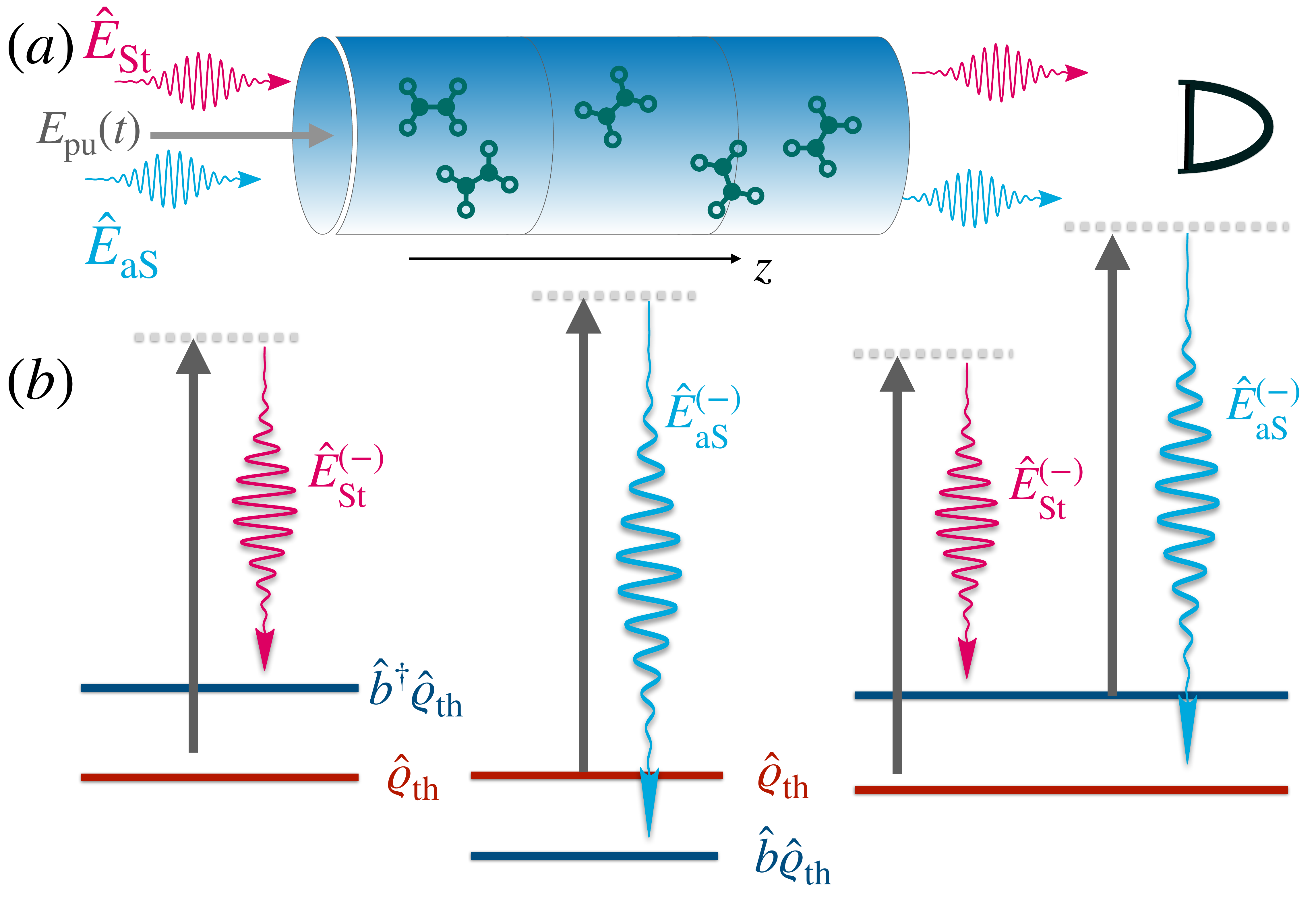}
    \caption{
    (a) Sketch of the setup considered here. We discuss Stokes-anti-Stokes-scattering in a collinear setup, where the sample is contained in a volume of length $L$.
    (b) Inelastic scattering processes included here: individual scattering processes involving only Stokes or anti-Stokes fields change the phonon populations. In addition, correlated events are mediated by the sample without changing its phonon population.
    %Properties of the emitted fields vs. the detuning and the effective scattering strength $\Gamma_{\mathrm{St-aS}}$.
    }
    \label{fig:sketch}
\end{figure}

In the following, we consider nonlinear wave-mixing processes generated locally through the interaction with point-like molecules, as described by Eq.~(\ref{eq:H_I}). These processes are sketched in Fig.~\ref{fig:sketch}(b) and include inelastic scattering events that generate Stokes and anti-Stokes photons, respectively, thereby heating or cooling the sample, as well as correlated Stokes--anti-Stokes scattering that produces quantum-correlated photon pairs. In this setting, the above molecular correlation functions become local as well.

The full derivation of the cumulants is presented in Appendix~\ref{app:derivation}. We retain only the leading non-vanishing cumulants and assume the phonons to be in a thermal state characterized by the mean phonon number $n_{\mathrm{th}}$. Consequently, the phonon dynamics are fully captured by their two-point correlation function. This constitutes an approximation that is sufficient for a phenomenological description of line broadening, but it does not, for example, distinguish between longitudinal and transverse relaxation times  (conventionally denoted $T_1$ and $T_2$, respectively), dynamical information that would be encoded in a four-point correlation function. Spatial integration gives rise to phase-matching effects, which we discuss in Appendix~\ref{sec:propagation-effects}.

%Combining Eqs.~\eqref{eq.K_St-aS_full}, \eqref{eq.K_St}, and \eqref{eq.K_aS}, we find the quantum map takes the form of a cascaded master equation~\cite{Gardiner1993, Carmichael1993}, which we can combine into a Lindblad form. It is convenient to define effective two-photon operators
Combining Eqs.~\eqref{eq.K_St-aS_full}, \eqref{eq.K_St}, and \eqref{eq.K_aS}, we find that the quantum map takes the form of a cascaded master equation~\cite{Gardiner1993, Carmichael1993}, which can be recast in Lindblad form. It is convenient to introduce effective two-photon operators
\begin{align} \label{eq:epsilon_St}
    \hat{\epsilon}^\dagger_{\mathrm{St}} (\omega_-) &= \int \frac{d\omega}{2\pi} E_{\mathrm{pu}} (\omega + \omega_-) \hat{E}^\dagger_{\mathrm{St}} (\omega)
\end{align}
and
\begin{align} \label{eq:epsilon_aS}
    \hat{\epsilon}^\dagger_{\mathrm{aS}} (\omega_-) &= \int \frac{d\omega}{2\pi} E_{\mathrm{pu}} (\omega - \omega_-) \hat{E}^\dagger_{\mathrm{aS}} (\omega),
\end{align}
which are given by the overlap integral between the classical pump field shifted by $\pm \omega_-$ and the emitted quantum fields. 
Upon normalization, these operators define Stokes and anti-Stokes broadband modes which depend on the bandwidth of the pump pulse. 
The full cumulant then takes the form
\begin{widetext}
% \mathcal{K}_{\mathrm{St-aS}} (\vec{r}) + \mathcal{K}_{\mathrm{St}} (\vec{r}) + \mathcal{K}_{\mathrm{aS}} (\vec{r}) =
\begin{align}
    &\mathcal{K}_{\mathrm{St-aS}} \hat{\varrho} = \int \frac{d\omega_-}{2\pi} \bigg\{ - i (2 n_{\mathrm{th}}+1) \left[ \hat{H}_{\mathrm{eff}}, \hat{\varrho} \right]  \notag \\
    &+ \gamma_{\mathrm{St-aS}} (n_{\mathrm{th}}+1) \bigg( \mathcal{D} [\hat{\epsilon}^\dagger_{\mathrm{St}}] \hat{\varrho} + \mathcal{D} [\hat{\epsilon}_{\mathrm{aS}}] \hat{\varrho} 
    - \frac{\eta (\omega_-)}{2} \left\{ \left[ \hat{\epsilon}^\dagger_{\mathrm{aS}}, \hat{\epsilon}^\dagger_{\mathrm{St}} \hat{\varrho}\right] + \left[\hat{\varrho} \hat{\epsilon}^\dagger_{\mathrm{aS}}, \hat{\epsilon}^\dagger_{\mathrm{St}} \right] + \left[ \hat{\epsilon}_{\mathrm{St}}, \hat{\epsilon}_{\mathrm{aS}} \hat{\varrho}\right] + \left[ \hat{\varrho} \hat{\epsilon}_{\mathrm{St}} ,  \hat{\epsilon}_{\mathrm{aS}}\right] \right\}
    \bigg)\notag\\
    &+ \gamma_{\mathrm{St-aS}} n_{\mathrm{th}} \bigg( \mathcal{D} [\hat{\epsilon}_{\mathrm{St}}] \hat{\varrho} + \mathcal{D} [\hat{\epsilon}^\dagger_{\mathrm{aS}}] \hat{\varrho} 
    - \frac{\eta (\omega_-)}{2} \left\{ \left[ \hat{\epsilon}^\dagger_{\mathrm{St}}, \hat{\epsilon}^\dagger_{\mathrm{aS}} \hat{\varrho}\right] + \left[ \hat{\varrho} \hat{\epsilon}^\dagger_{\mathrm{St}}, \hat{\epsilon}^\dagger_{\mathrm{aS}}\right] + \left[ \hat{\epsilon}_{\mathrm{aS}}, \hat{\epsilon}_{\mathrm{St}} \hat{\varrho}\right] + \left[ \hat{\varrho} \hat{\epsilon}_{\mathrm{aS}}, \hat{\epsilon}_{\mathrm{St}} \right] \right\}
    \bigg) \bigg\}, \label{eq.full-cumulant}
\end{align}
\end{widetext}
%\textcolor{red}{check last line}
where $\mathcal{D}[x] \hat{\varrho} = x \hat{\varrho} x^\dagger - \{ x^\dagger x, \hat{\varrho} \} / 2$ denotes a Lindblad dissipator. The evolution contains an effective Hamiltonian $\hat{H}_{\mathrm{eff}}$ which we describe below. 
The second line collects dissipative processes scaling $\sim n_{\mathrm{th}}+1$ that can occur spontaneously. These include Stokes gain described by $\mathcal{D} [\hat{\epsilon}^\dagger_{\mathrm{St}}]$ and anti-Stokes loss, $\mathcal{D} [\hat{\epsilon}_{\mathrm{aS}}]$. 
The final term accounts for the cascaded,  correlated emission of Stokes-anti-Stokes photon pairs (or their absorption). 
Stokes emission (anti-Stokes absorption) triggers anti-Stokes emission (Stokes absorption). 
The function $\eta (\omega_-)$ accounts for phasematching restrictions due to propagation effects which may reduce the collective dynamics terms in the final curly brackets, and therefore $|\eta (\omega_-)|\leq 1$.  
The third line contains the inverse processes, where the role of Stokes and anti-Stokes fields are exchanged. These consequently scale like $\sim n_{\mathrm{th}}$.  
Finally, we have defined the effective Hamiltonian
\begin{widetext}
\begin{align} \label{eq:H_eff}
    \hat{H}_{\mathrm{eff}} = h_{\mathrm{St-aS}} ( \hat{\epsilon}_{\mathrm{St}} \hat{\epsilon}_{\mathrm{St}}^\dagger + \hat{\epsilon}^\dagger_{\mathrm{aS}} \hat{\epsilon}_{\mathrm{aS}} + \eta (\omega_-) \big( \hat{\epsilon}_{\mathrm{St}} \hat{\epsilon}_{\mathrm{aS}} + \hat{\epsilon}^\dagger_{\mathrm{St}} \hat{\epsilon}^\dagger_{\mathrm{aS}} \big)).
\end{align}
\end{widetext}
Eq.~(\ref{eq:H_eff}) includes a two-mode squeezing Hamiltonian, as well as terms proportional to the Stokes and anti-Stokes intensity, respectively. 
The two-mode squeezing describes the coherent generation of entangled Stokes-anti-Stokes pairs, or their absorption by the sample. 
The latter terms account for frequency-dependent dispersion induced by the sample, and activated by the pump pulse. 
%a linear dispersion induced by the sample. 
%\textcolor{red}{no, doesnt quite work this way, only for cw..}

The effective Hamiltonian prefactor $h_{\mathrm{St-aS}}$ and the decay parameter $\gamma_{\mathrm{St-aS}}$ are directly connected to the molecular ensemble's Raman response,
\begin{align}
    R_{\mathrm{St-aS}} &= - i  N \frac{ |\alpha_{\mathrm{R}}|^2 }{\hbar^2} \frac{1}{\omega_- - \omega_{\mathrm{ph}} + i \gamma} \notag \\
    &= - i h_{\mathrm{St-aS}} - \gamma_{\mathrm{St-aS}} /2, \label{eq:R_St-aS}
\end{align}
where $N$ is the number of molecules in the sample, $\gamma$ is the effective broadening of the phonon line, and $\omega_-$ denotes the detuning of the two-photon interaction in Eq.~(\ref{eq:H_I}) from the phonon resonance $\omega_{\mathrm{ph}}$. Specifically, we have
\begin{align}
    h_{\mathrm{St-aS}} (\omega_-) &= \frac{ N |\alpha_{\mathrm{R}}|^2}{\hbar^2} \frac{\omega_- - \omega_{\mathrm{ph}}}{(\omega_- - \omega_{\mathrm{ph}})^2 + \gamma^2}
\end{align}
and
\begin{align}
    \gamma_{\mathrm{St-aS}} (\omega_-) &= \frac{N |\alpha_{\mathrm{R}}|^2}{\hbar^2} \frac{ 2\gamma}{(\omega_- - \omega_{\mathrm{ph}})^2 + \gamma^2}.
\end{align}
Note that both $h_{\mathrm{St-aS}}$ and $\gamma_{\mathrm{St-aS}}$ must not be interpreted as energies or rates, since the full cumulant~(\ref{eq.full-cumulant}) is dimensionless.  
When $|\omega_- - \omega_{\mathrm{ph}}| \gg \gamma$, the effective Hamiltonian dominates and the scattering looks like a broadband squeezing operation. This is the well-established parametric regime of quantum optics, where an effective Hamiltonian description is possible. 
But near resonance, $|\omega_- - \omega_{\mathrm{ph}}| < \gamma$, when real phonons are excited or destroyed in the sample, energy can be exchanged between field and sample. This renders the associated photonic dynamics (partially) incoherent.

When propagation effects can be neglected,
i.e. $\eta (\omega_-) = 1$, %and the excitation is balanced, i.e. $\gamma_{\mathrm{St}} = \gamma_{\mathrm{aS}}$,  
we can cast the dissipator into Lindblad form with a single collective decay operator, 
\begin{align} \label{eq.c-Lindbladian}
    &\mathcal{K}_{\mathrm{St-aS}} \hat{\varrho} = \int \frac{d\omega_-}{2\pi} \bigg\{ - i(2 n_{\mathrm{th}}+1) \left[ \hat{H}_{\mathrm{eff}}, \hat{\varrho} \right] \notag \\
    &+ 2\gamma_{\mathrm{St-aS}} (n_{\mathrm{th}}+1) \mathcal{D} [\hat{c}] \hat{\varrho} +2\gamma_{\mathrm{St-aS}}  n_{\mathrm{th}} \mathcal{D} [\hat{c}^\dagger] \hat{\varrho} \bigg\}, 
\end{align}
%\textcolor{red}{check last line}
where $\hat{c} = (\hat{\epsilon}_{\mathrm{aS}} + \hat{\epsilon}^\dagger_{\mathrm{St}} )/\sqrt{2}$ is a collective non-Hermitian operator.  Propagation effects in a deep sample restrict the coupled Stokes-anti-Stokes contributions. Hence, we obtain dynamics with more pronounced Stokes (and anti-Stokes) gain (loss), respectively. Apart from this, the dynamics remains qualitatively similar. 
%\textcolor{red}{In the presence of propagation effects, collective contribution reduced, hence, weg get gain + loss in Stokes, aS, respectively}

%The nonreciprocity depends on the phonon temperature. In the high-temperature limit $n_{\mathrm{th}} \gg 1$, we may neglect the spontaneous contribution in the second line of Eq.~(\ref{eq.full-cumulant}). The \textcolor{red}{to be added}

We note that Eq.~(\ref{eq.full-cumulant}) depends implicitly on the pump bandwidth through Eqs.~(\ref{eq:epsilon_St}) and (\ref{eq:epsilon_aS}). Hence, Eq.~(\ref{eq.full-cumulant}) describes, in general, a highly multimode evolution that mixes and entangles Stokes and anti-Stokes photons in their frequency degrees of freedom. 
We only recover a two-mode description in certain limiting cases.

%\subsubsection{Non-Hermitian dynamics}
%The effective non-Hermitian Hamiltonian reads
%\begin{align}
%    &H_{non-H} \notag \\
%    &= (2n_{\mathrm{th}}+1) \hat{H}_{\mathrm{eff}} - \frac{i}{2} \left( (n_{\mathrm{th}}+1) \hat{c}^\dagger \hat{c} + n_{\mathrm{th}} \hat{c}\hat{c}^\dagger \right) \notag \\
%    &= \frac{2n_{\mathrm{th}}+1}{\Delta + i \gamma} \epsilon_{\mathrm{St}} \epsilon_{\mathrm{aS}} + \epsilon^\dagger_{\mathrm{St}} \epsilon^\dagger_{\mathrm{aS}} + \epsilon_{\mathrm{St}} \epsilon_{\mathrm{St}}^\dagger + \epsilon^\dagger_{\mathrm{aS}} \epsilon_{\mathrm{aS}}
%\end{align}
%In the present case, this does not show exceptional points, but when we have two Raman lines, there will be exceptional points when the two resonances merge as a function of $\gamma$.

\subsection{cw pump}
When the scattering is driven by a cw pump, we write the corresponding field amplitude as %(setting the sample at the origin $z = 0$)
\begin{align} \label{eq:cw-pump}
    E_{\mathrm{pu}}^{\mathrm{(cw)}} (t) &= E_{\mathrm{pu}} e^{i (k_{\mathrm{pu}} z -  \omega_{\mathrm{pu}} t)}.
\end{align}
The two-photon operators $\epsilon_{\mathrm{St}}$ and $\epsilon_{\mathrm{aS}}$ then become proportional to the shifted photon annihilation operators, i.e. $\epsilon^{\mathrm{(cw)}}_{\mathrm{St}} (\omega_-)\propto \hat{a}_{\mathrm{St}} (\omega_{\mathrm{pu}} - \omega_-)$
and 
$\epsilon^{\mathrm{(cw)}}_{\mathrm{aS}} (\omega_-)\propto \hat{a}_{\mathrm{aS}} (\omega_{\mathrm{pu}} + \omega_-)$.
Thus, Eq.~(\ref{eq.full-cumulant}) becomes a master equation coupling frequency pairs $\omega \pm \omega_-$. In this sense, it looks similar to a broadband squeezing operation stimulated by a cw pump, with the important distinction that the resonant phonon excitation renders the field dynamics partially dissipative. 
Phase matching or spectral filtering may be used to select a specific pair or frequency range. 
%Consequently, we may 
In the absence of phase matching restrictions, we pleace the sample at $z = 0$ and
describe the cw pump regime with Eq.~(\ref{eq.full-cumulant}), when we replace $\epsilon_{\mathrm{St/aS}} (\omega_-)\rightarrow \hat{a}_{\mathrm{St/aS}} (\omega_{\mathrm{pu}} \pm \omega_-)$, and rescale the Hamiltonian and decay parameters $(h_{\mathrm{St-aS}}, \gamma_{\mathrm{St-aS}}) \rightarrow E_{\mathrm{pu}}^2 E_{\mathrm{vac}}^2 (h_{\mathrm{St-aS}}, \gamma_{\mathrm{St-aS}})$.

\begin{figure}
    \centering
    \includegraphics[width=\linewidth]{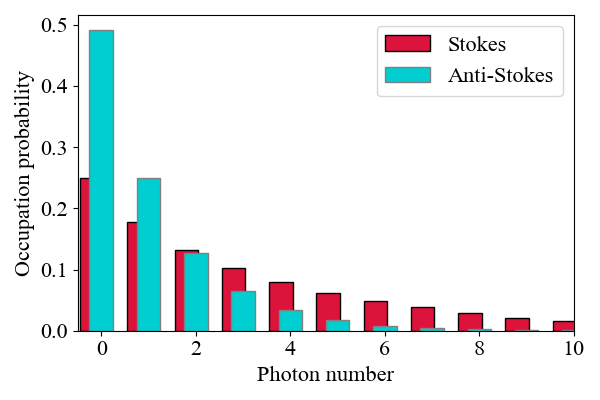}%{photon-statistics3.png}
    \caption{Photon distribution of Stokes and anti-Stokes fields for resonant excitation (i.e. $\Delta = 0$), zero phonon population $n_{\mathrm{th}} = 0.0$, and conversion strength $\Gamma_{\mathrm{cw}} = 1.0$. }
    \label{fig:statistics}
\end{figure}

\subsection{ultrafast pump}

In the opposite limit, we consider driving by an ultrafast pulse which we treat in the impulsive limit, i.e. we assume that the pump duration is much shorter than the timescale of the phonon dynamics (for instance, a $100$~fs pump is a 1000-times faster than phonon dynamics on the scale of $10$~ps).
We further assume that the pump bandwidth does not directly overlap with the Stokes and anti-Stokes fields, % This allows us to 
such that we may still use the Hamiltonian~(\ref{eq:H_I}) \footnote{If this condition was violated, self-interactions of the fields would have to be taken into account. This can be done with the present formalism, but requires the inclusion of additional Feynman diagrams and complicates the analysis.}. 
%In the impulsive excitation regime, 
We thus write the pump field as
\begin{align} \label{eq:impulsive-pump}
    E_{\mathrm{pu}} (t) &= E_{\mathrm{pu}} \delta (t - t_0) e^{- i \omega_{\mathrm{pu}} t} / f_{\mathrm{rep}},
\end{align}
where $f_{\mathrm{rep}}$ is the repetition rate of the pulsed laser, introduced here for consistency, such that $E_{\mathrm{pu}}$ has the same units as Eq.~(\ref{eq:cw-pump}). 
It is convenient to evaluate the Feynman diagrams directly in time domain (see Appendix~\ref{app:impulsive-excitation}), and we arrive at, neglecting propagation effects, 
\begin{align}
    \mathcal{K}_{\mathrm{pulse}} \hat{\varrho} &= \gamma_{\mathrm{pulse}} (n_{\mathrm{th}} + 1) \mathcal{D} [\hat{c}'] \hat{\varrho} + \gamma_{\mathrm{pulse}} n_{\mathrm{th}} \mathcal{D} [\hat{c}'^\dagger] \hat{\varrho}, 
\end{align}
with $\hat{c}' = \hat{a}_{\mathrm{aS}} (t_0) + \hat{a}^\dagger_{\mathrm{St}} (t_0)$, and 
\begin{align} \label{eq:gamma_pulse}
    \gamma_{\mathrm{pulse}} &= \frac{N | \alpha_{\mathrm{R}}|^2 E_{\mathrm{pu}}^2 E_{\mathrm{vac}}^2}{\hbar^2 f^2_{\mathrm{rep}}}.
\end{align}
Hence, we obtain purely dissipative dynamics with the impulsive emission of broadband Stokes and anti-Stokes photons. The effective Hamiltonian vanishes due to the symmetry of the Raman response. 
%and we can directly apply the insights from cw excitation. %we have already gained in the previous section for resonant excitation. 

%This changes in Fig.~\ref{fig:sketch}(c), where we show a strong scattering regime with $\Gamma_{\mathrm{Raman}} = 0.3$. 

\begin{figure}[t!]
    \centering
    \includegraphics[width=\linewidth]{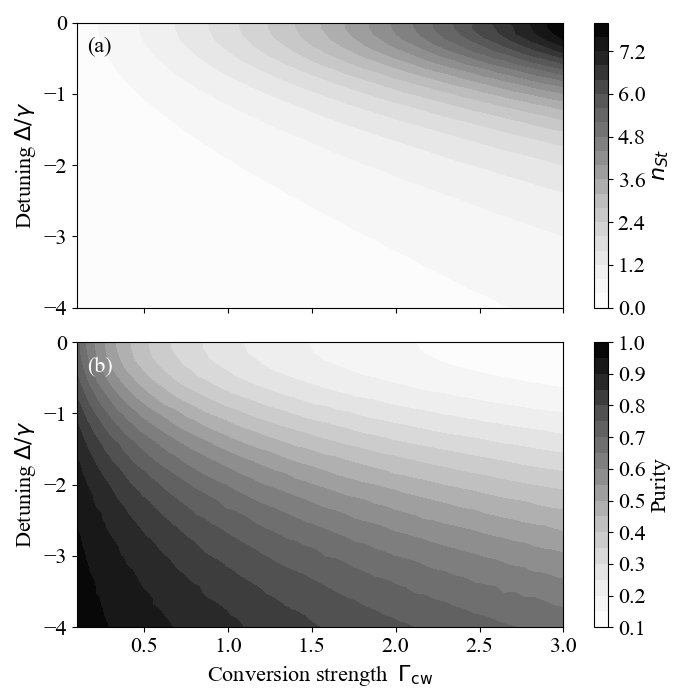}
    \caption{
    (a) The mean Stokes photon umber $\langle \hat{n}_{\mathrm{St}} \rangle$, for cw Raman scattering, is plotted vs. the detuning $\Delta$ and the cw conversion strength $\Gamma_{cw}$. 
    (b) same a (a) for the purity of the output photonic state. 
    In both plots, we consider a sample with $n_{\mathrm{th}} = 0$.}
    \label{fig:cw-contours}
\end{figure}

\begin{figure}[t]
    \centering
    \includegraphics[width=\linewidth]{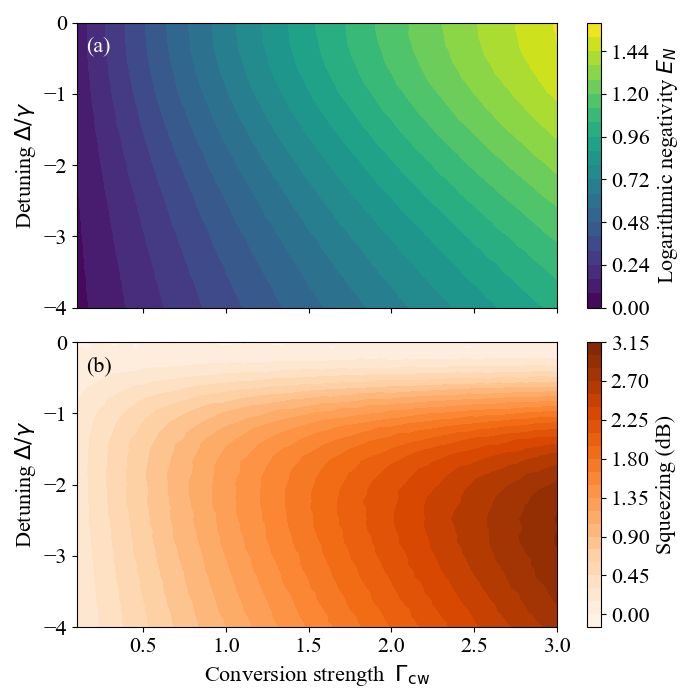}
    \caption{
    (a) The logarithmic negativity $E_N$~(\ref{eq:E_N}), for cw Raman scattering, is plotted vs. the detuning $\Delta$ and the cw conversion strength $\Gamma_{\mathrm{cw}}$.  
    (b) same as (a) for the degree of squeezing of the output photonic state. 
    In both plots, we consider a sample with $n_{\mathrm{th}} = 0$.
    }
    \label{fig:cw-contours2}
\end{figure}

\section{Simulations}
\label{sec:simulations}

We first focus on Raman scattering induced by a cw pump~(\ref{eq:cw-pump}) and numerically propagate Eq.~(\ref{eq:auxilliary}) for two modes, described by the dimensionless operators $\hat{a}_{\mathrm{St}} = \hat{a}_{\mathrm{St}} (\omega) \Delta\omega^{1/2}$ and $\hat{a}_{\mathrm{aS}} = \hat{a}_{\mathrm{aS}} (\omega) \Delta\omega^{1/2}$. Here, $\Delta \omega$ denotes a small frequency interval. The Raman conversion strength is quantified by $\Gamma_{\mathrm{cw}} = N |\alpha_{\mathrm{R}}|^2 E_{\mathrm{pu}}^2 E_{\mathrm{vac}}^2/ \hbar^2\gamma$, such that $h_{\mathrm{St-aS}} = \Gamma_{\mathrm{cw}} \Delta \gamma / (\Delta^2 + \gamma^2) $ and $\gamma_{\mathrm{St-aS}} = \Gamma_{\mathrm{cw}} \gamma^2 / (\Delta^2 + \gamma^2)$. 

An example of the photon distributions generated by resonant Raman scattering is shown in Fig.~\ref{fig:statistics}. Evidently, the number of Stokes photons produced greatly exceeds that of anti-Stokes photons at zero temperature, since the latter can only be generated through correlated Stokes-anti-Stokes scattering events. As consequence, the emitted fields must be in a mixed state. 

This is further explored in Fig.~\ref{fig:cw-contours}(a), where we present the mean Stokes photon number, $\langle \hat{a}^\dagger_{\mathrm{St}} \hat{a}_{\mathrm{St}}\rangle$, for a single high-frequency phonon mode with $n_{\mathrm{th}} = 0$ as a function of the detuning $\Delta = \omega_{-} - \omega_{\mathrm{ph}}$ (measured in units of $\gamma$) and scattering strength $\Gamma_{\mathrm{cw}}$. It clearly shows the resonant enhancement of the emission intensity. Simultaneously, the purity of the photonic density matrix, shown in the bottom panel, is reduced by resonant scattering and decreases with increasing photon number. 
\\

\subsection{Entanglement and correlations}

%as a function of the detuning $\Delta = \omega_{-} - \omega_{\mathrm{ph}}$ (measured in units of $\gamma$) and the thermal phonon number $n_{\mathrm{th}}$.  

%In Fig.~\ref{fig:sketch}(b), we present simulations in the weak scattering regime, $\Gamma_{\mathrm{Raman}} = 10^{-3}$, where the mean photon number in each beam is much smaller than 1. We characterize the amount of distillable entanglement with the logarithmic negativity in the left panel, showing that in this weak scattering regime the largest amount of entanglement is generated on resonance. 
 
Yet despite this reduction of the state's purity, the distillable entanglement appears to increase with the conversion strength $\Gamma_{\mathrm{cw}}$.  
We trace this behaviour with the logarithmic negativity~\cite{Plenio2005}, which upper-bounds the distillable entanglement in the emitted fields, defined by
\begin{align} \label{eq:E_N}
    E_N &= \log_2 ||\hat{\varrho}_{\mathrm{out}}^{T}||_1,
\end{align}
where $\varrho^{T}$ denotes the partial transpose of output density matrix, and $||X||_1 = \text{tr} \sqrt{ \hat{X}^\dagger \hat{X} }$ the trace norm. 
%resonance does not directly translate into an enhancement of the generated entanglement, or of squeezing: 
%present simulations of the logarithmic negativity which we chose to characterize the amount of distillable entanglement in the generated light field. It is shown for a high-frequency phonon mode with $n_{\mathrm{th}} = 0$ vs. detuning $\Delta = \omega_{-} - \omega_{\mathrm{ph}}$ (measured in units of $\gamma$) and scattering strength $\Gamma_{\mathrm{St-aS}}$. 
It is simulated in Fig.~\ref{fig:cw-contours2}(a) and broadly follows the trend of the mean Stokes number in Fig.~\ref{fig:cw-contours})(a). % and increases in line with the mean photon number. 
%depicts the logarithmic negativity of the emitted light 
%which we chose to characterize the amount of distillable entanglement.  
%At weak conversion, when $\Gamma_{\mathrm{cw}} \lesssim 2$, the largest logarithmic negativity is obtained on resonance $\Delta = 0$, i.e. when the output intensity is maximized. This can be understood as akin to the isolated pair regime in parametric downconversion, where the emitted fields are predominantly composed of isolated, entangled pairs, with the important distinction here that the output state is never pure. However, the projection of this state onto the subspace spanned by one photon in the Stokes and anti-Stokes sector, respectively, is approximately pure, i.e.
This can be rationalized by the presence of entangled pairs in the emitted light.
In the full multimode description~(\ref{eq.full-cumulant}), we can identify terms that generate photon pairs of the form
\begin{widetext}
\begin{align} \label{eq:psi_ent}
    \vert \psi_{\mathrm{ent}} \rangle &= \int \frac{d\omega_-}{2\pi} \Phi (\omega_-)\hat{a}^\dagger_{\mathrm{aS}}(\omega_{\mathrm{pu}}+ \omega_-) \hat{a}^\dagger_{\mathrm{St}}(\omega_{\mathrm{pu}} - \omega_-) \vert 0 \rangle,
\end{align}
\end{widetext}
and
\begin{align} \label{eq:Phi}
    \Phi (\omega_-) &=\frac{1}{\sqrt{\mathcal{N}}} \frac{ \text{sinc} \left( \frac{\Delta k L}{2} \right) e^{i \Delta k L /2} }{\omega_- - \omega_{\mathrm{ph}} + i \gamma},
\end{align}
where $\mathcal{N}$ is a normalization factor and the sinc-function arises from phase matching (see Appendix~\ref{sec:propagation-effects}). 
In the weak scattering regime, where $\langle \hat{a}^\dagger_{\mathrm{St}} \hat{a}_{\mathrm{St}} \rangle \ll 1$, these pairs dominate the two-photon sector of the emitted light, i.e. we can write %(up to $\mathcal{O} (\Gamma_{\mathrm{St-aS}}^2)$)
\begin{align}
   \hat{P}_{1\mathrm{St}, 1\mathrm{aS}} \hat{\varrho}_{\mathrm{out}}\hat{P}_{1\mathrm{St}, 1\mathrm{aS}} = \vert \psi_{\mathrm{ent}} \rangle\langle\psi_{\mathrm{ent}} \vert + \mathcal{O} (\Gamma_{\mathrm{cw}}^2),
\end{align}
where $\hat{P}_{1\mathrm{St}, 1\mathrm{aS}}$ denotes the projector onto the subspace containing one Stokes and one anti-Stokes photon each. 
This can be understood as akin to the isolated pair regime in parametric downconversion, where the emitted fields are predominantly composed of isolated, entangled pairs, with the important distinction here that the output state is never pure. 
Here, however, these pairs are mixed with individual Stokes or anti-Stokes photons from different scattering events. 
%However, the projection of this state onto the subspace spanned by one photon in the Stokes and anti-Stokes sector, respectively, is approximately pure. 
Still, at this level of isolated pairs, if one carries out a photon coincidence measurement~\cite{Saraiva2017, Anderson2018}, the signal is indistinguishable from a signal produced by entangled photon pairs, even though the 'full' quantum state is always mixed, as we will explore below. 
In this situation, one may interpret Eq.~(\ref{eq:Phi}) as the two-photon wavefunction or joint spectral amplitude of the entangled pair. 
Due to the two-photon resonance, $\omega_{\mathrm{pu}} - \omega_{\mathrm{St}} = \omega_{\mathrm{ph}}$, this wavefunction inherits a strong phase dependence which must be accounted for to accurately extract its spectral entanglement in a multimode regime. 

%When the number of emitted photons increases, it becomes beneficial to activate the Hamiltonian contribution in Eq.~(\ref{eq.full-cumulant}) which contains a two-mode squeezing Hamiltonian. Consequently, to maximize the logarithmic negativity for $\Gamma_{cw} \gtrsim 2$, a finite detuning becomes optimal, as indicated in Fig.~\ref{fig:cw-contours}(b). 
Still, important differences emerge from the dissipative nature of the resonant downconversion. This is manifest in the absence of squeezing below the vacuum limit. 
It is shown in the bottom panel of Fig.~\ref{fig:cw-contours2}, where we plot the squeezing factor
\begin{align}
    S &\equiv 10 \log_{10} \left( \frac{ \langle \Delta\hat{q}^2 \rangle_{\mathrm{vac}} }{ \langle \Delta\hat{q}^2_{\mathrm{min}} \rangle_{\mathrm{out}} } \right),
\end{align}
where $\Delta\hat{q}^2_{\mathrm{min}} $ denotes the minimal quadrature fluctuation evaluated with the output state~(\ref{eq:rho_out-def}).
To generate squeezing below the vacuum noise limit, it is strictly necessary to keep a finite detuning - just like in four-wave mixing in hot atomic vapours~\cite{Yang2020, DorfmanPNAS}. 
The resonantly emitted field, despite being entangled, only shows %negative squeezing, i.e. 
enhanced fluctuations. %compared to the vacuum. 
%Due to its reduced purity, the state is not Fourier-limited. 
Squeezing below the vacuum can only be generated when the Hamiltonian in Eq.~(\ref{eq.full-cumulant}) dominates the photonic dynamics. 
Moreover, as becomes evident from the bottom panel of Fig.~\ref{fig:cw-contours2}, the detuning necessary to optimize the degree of squeezing at a given pump intensity (and hence conversion strength) increases approximately linearly with said intensity. 

\subsection{Resonant scattering}

We now focus on resonant scattering, and investigate the properties of the two sideband fields further. 
A natural question is how the phonon temperature affects the emitted radiation. 
In Eq.~(\ref{eq.full-cumulant}), we see that %, by virtue of the bosonic nature of the phonons, 
the effective coupling strength for fixed geometry and pump intensity increases linearly with $n_{\mathrm{th}}$. 
Hence, the effective spontaneous Raman cross section increases~\cite{Batignani2024}. At the same time, however, the effective rates in the second and third lines of Eq.~(\ref{eq.full-cumulant}) become more similar, such that the effective dynamics could becomes more symmetric. 
%and 
It is therefore interesting how this will affect quantum correlations of the emitted Raman radiation. 

\begin{figure}
    \centering
    \includegraphics[width=0.99\linewidth]{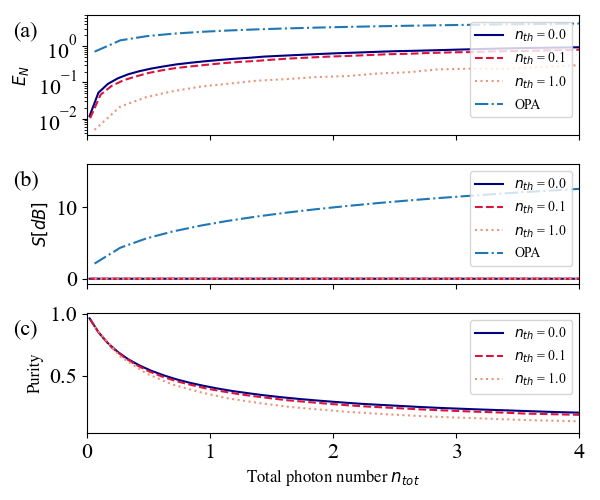}
    \caption{Logarithmic negativity~(\ref{eq:E_N}) [top panel], squeezing of collective Stokes and anti-Stokes quadratures [center panel] and purity [bottom panel] are plotted vs the total photon number. In each case, the corresponding simulations are compared to a two-mode squeezed state~(\ref{eq:tms-state}), labelled OPA. }
    \label{fig:phonon-temperature}
\end{figure}

To this end, we compare the output state~(\ref{eq:rho_out-def}) to an idealized two-mode squeezed state 
\begin{align}
    \vert \psi_{\mathrm{OPA}} \rangle &= \frac{1}{\cosh (r)}\sum_{n= 0}^\infty \tanh^n (r) \vert n \rangle_{\mathrm{St}} \otimes \vert n\rangle_{\mathrm{aS}}, \label{eq:tms-state}
\end{align}
where $r$ is the squeezing parameter, such that $n_{\mathrm{St}}^{\mathrm{(OPA)}} = n_{\mathrm{aS}}^{\mathrm{(OPA)}} = \sinh^2 r$.
This state gives rise to the logarithmic negativity~\cite{Giedke2003}
\begin{align}
    E_N^{\mathrm{(OPA)}} &= 2 r \log_2 e,
\end{align}
and two-mode squeezing (in dB)
\begin{align}
    S_{\mathrm{OPA}} &= - 10 \log_{10} e^{-2r}.
\end{align}
In Fig.~\ref{fig:phonon-temperature}, we depict the evolution of the logarithmic negativity, the squeezing and the output state purity vs. the mean total photon number $n_{\mathrm{tot}} = \langle \hat{a}^\dagger_{\mathrm{St}} \hat{a}_{\mathrm{St}} + \hat{a}^\dagger_{\mathrm{aS}} \hat{a}_{\mathrm{aS}} \rangle$. 
Even at $n_{\mathrm{th}} = 0$, the generated entanglement remains about two orders of magnitude below the level of a squeezed vacuum~(\ref{eq:tms-state}) with the same mean total photon number due to the prevalence of unpaired photons. It further decreases rapidly with increasing phonon temperature, despite the larger apparent Raman cross sections available for correlated scattering events.  
This reduction is not caused by the choice to plot it as a function of the total photon number. It also persists when we compare it at identical Raman conversion strengths $\Gamma_{\mathrm{cw}}$. 
Rather, the increased phonon temperature makes the emitted light more noisy and thus reduces the relative impact of quantum correlations. 
This can be further rationalised from the purity of the emitted light, which decreases slightly with increasing $n_{\mathrm{th}}$, as shown in Fig.~\ref{fig:phonon-temperature}(c).
Squeezing, as discussed previously, is never generated by resonant Raman scattering at any phonon temperature.

\begin{figure}
    \centering
    \includegraphics[width=\linewidth]{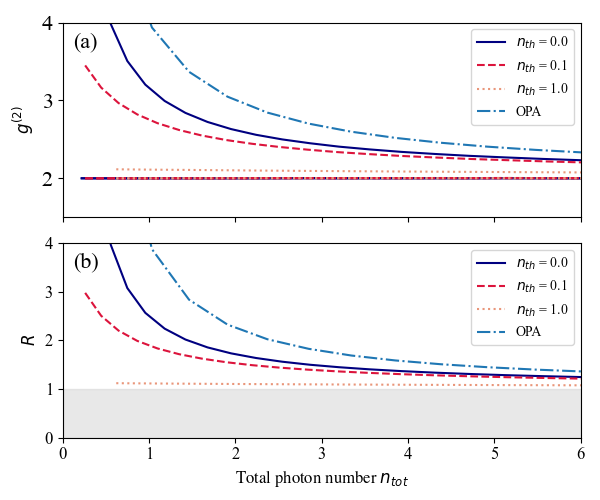}
    \caption{(a) $g^{(2)}$ functions of the light field, Eq.~(\ref{eq:g^2_StaS}) and (\ref{eq:g^2_St}) are shown as functions of the mean total photon number for different phonon temperatures. The constant plots at $g^{(2)} = 2$ correspond to the intramode correlations~(\ref{eq:g^2_St}). 
    (b) From the above correlation functions, we construct the nonclassicality witness~(\ref{eq:R-factor}). The grey shaded area indicates regimes where the nonclassicality of the radiation cannot be ascertained. In each case, the corresponding simulations are compared to a two-mode squeezed state~(\ref{eq:tms-state}), labelled OPA.}
    \label{fig:nonclassicality}
\end{figure}

These trends are also reflected in the nonclassicality of the emitted light fields. 
To this end, we simulate the cross correlation function (at equal times)
\begin{align} \label{eq:g^2_StaS}
    g^{(2)}_{\mathrm{St-aS}} &= \frac{ \langle \hat{a}_{\mathrm{St}}^\dagger \hat{a}_{\mathrm{aS}}^\dagger \hat{a}_{\mathrm{aS}} \hat{a}_{\mathrm{St}} \rangle }{ \langle \hat{a}^\dagger_{\mathrm{St}} \hat{a}_{\mathrm{St}}  \rangle \langle \hat{a}^\dagger_{\mathrm{aS}} \hat{a}_{\mathrm{aS}}  \rangle }
\end{align}
and the intramode photon correlations 
\begin{align} \label{eq:g^2_St}
    g^{(2)}_{j} &= \frac{ \langle \left( \hat{a}^\dagger_{j}\right)^2 \hat{a}_{j}^2 \rangle }{ \langle \hat{a}^\dagger_{j} \hat{a}_{j}  \rangle^2 },
\end{align}
where $j =$ St, aS. For the two-mode squeezed vacuum state~(\ref{eq:tms-state}), we find $g^{(2)}_{\mathrm{St, OPA}} = g^{(2)}_{\mathrm{aS, OPA}} = 2$. These chaotic fluctuations of the individual fields are reproduced in Raman scattering, as can be seen in the top panel of Fig.~\ref{fig:nonclassicality}. 
The cross correlations, which evaluate for $\vert \psi_{\mathrm{OPA}} \rangle$ to $g^{(2)}_{\mathrm{St-aS, OPA}} = 2 (1 + 1 / n_{\mathrm{tot}})$, are reduced substantially in resonant Raman scattering and decay more quickly to $2$ at large photon number. 

%and similarly for the anti-Stokes mode. 
These quantities can be combined to the nonclassicality witness~\cite{Saraiva2017, Anderson2018}
\begin{align} \label{eq:R-factor}
R &= \frac{ \left( g^{(2)}_{\mathrm{St-aS}} \right)^2 }{ g^{(2)}_{\mathrm{St}} g^{(2)}_{\mathrm{aS}} }.
\end{align}
A factor $R > 1$ indicates a nonclassical radiation source. 
This is shown in the bottom panel of Fig.~\ref{fig:nonclassicality}. For $\vert \psi_{\mathrm{OPA}} \rangle$, we find $R_{\mathrm{OPA}} = (1 + 1 / n_{\mathrm{tot}})^2 $. Hence, at large photon numbers, it becomes increasingly difficult to certify the nonclassicality of the Raman scattered light. 
The Raman correlation are more fragile, and they are further degraded by increasing thermal phonon occupation. Yet they show the same qualitative behaviour as $\vert \psi_{\mathrm{OPA}} \rangle$. 
This demonstrates that, as mentioned previously in the discussion of Eq.~(\ref{eq:psi_ent}), as far as photon correlations are concerned, correlations in Raman scattering and in parametric downconversion do look remarkably similar~\cite{Saraiva2017, Anderson2018}.

\subsection{Nonreciprocal photon amplification}

\begin{figure}
    \centering
    \includegraphics[width=\linewidth]{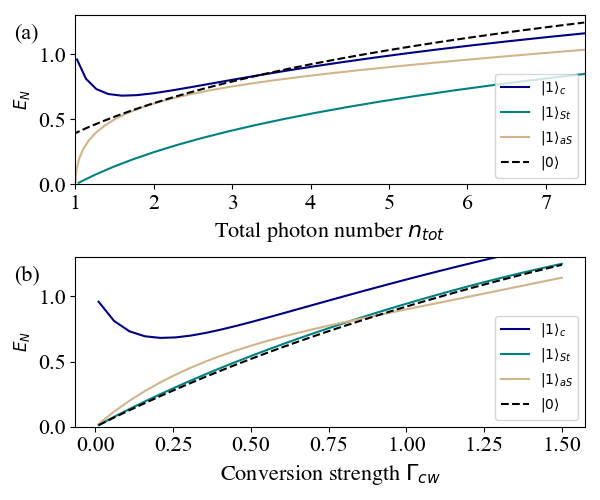}
    \caption{
    (a) Logarithmic negativity $E_N$~(\ref{eq:E_N}) of the Raman scattered fields is shown as a function of the mean total photon number $n_{\mathrm{tot}}$. The fields are seeded by single photons initialised in Eq.~(\ref{eq:1_c}), or in the Stokes or anti-Stokes modes, respectively. The vacuum result is shown as dashed, black line for comparison.  We set $n_{\mathrm{th}} = 0$.
     (b) Same data as in top panel is shown as a function of the Raman conversion strength $\Gamma_{\mathrm{cw}}$. }
    \label{fig:seed-photons}
\end{figure}

\begin{figure}
    \centering
    \includegraphics[width=\linewidth]{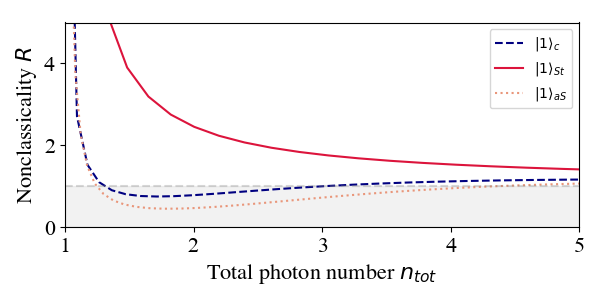}
    \caption{The nonclassicality witness $R$, Eq.~(\ref{eq:R-factor}), is shown as a function of the mean total photon number $n_{\mathrm{tot}}$. The fields are seeded by single photons initialised in either mode 'c', or in the Stokes or anti-Stokes modes, respectively. The vacuum result is shown as dashed, black line for comparison. We set $n_{\mathrm{th}} = 0$.}
    \label{fig:seeded-nonclassicality}
\end{figure}

\subsubsection{Seeded entanglement and photon correlations}
There are, however, also crucial differences. These emerge from the nonreciprocal nature of Raman scattering: at zero temperature, the emission of a Stokes photon may trigger the emission of an anti-Stokes photon, but not vice versa. 
This becomes apparent, for instance, when we seed the Raman sidebands: 
In Fig.~\ref{fig:seed-photons}, we plot the evolution of the logarithmic negativity $E_N$ as a function of the mean total photon number $n_{\mathrm{tot}}$, when the Raman scattering is stimulated by single photon states. For comparison, we also plot $E_N$ of the spontaneously generated signal as a dashed, black line. 

Seeding either the Stokes or anti-Stokes fields does not enhance $E_N$ relative to the vacuum state with the same mean total photon number. 
The anti-Stokes field can, however, almost saturate the logarithmic negativity of the spontaneous signal with $n_{\mathrm{tot}} \simeq 2$. 
This could point towards interesting way of amplifying entanglement in weak scattering situations where the spontaneous signal that can be generated is limited to small photon numbers. 
Overall, seeding the anti-Stokes field appears more efficient than seeding the Stokes, since the latter also stimulates uncorrelated Stokes emission. This puts Stokes seeding at a disadvantage when comparing entanglement at a fixed mean total photon number. 

We may instead want to optimise the entanglement content at a given pump intensity. 
In the bottom panel of Fig.~\ref{fig:seed-photons}, we depict the same simulations as a function of the Raman conversion strength $\Gamma_{\mathrm{cw}}$, which effectively amounts to a plot vs the pump intensity as the only experimentally controllable parameter in $\Gamma_{\mathrm{cw}}$. 
In this presentation, seeding the anti-Stokes field may boost entanglement only for $\Gamma_{\mathrm{cw}} \lesssim 0.75$. Seeding the Stokes field appears to always yields a marginal benefit. 

In principle, we can do better by seeding with superposition states. We define
\begin{align} \label{eq:1_c}
    \vert 1 \rangle_c &= \frac{1}{\sqrt{2}}(\hat{a}^\dagger_{\mathrm{St}} + \hat{a}^\dagger_{\mathrm{aS}}) \vert 0 \rangle, 
\end{align}
which is an entangled state of the two modes, yielding $E_N (\vert 1 \rangle_c) = 1$. Such a state could be generated through spectral filtering of a broadband single-photon pulse.
As can be seen in Fig.~\ref{fig:seed-photons}, the transmission through the Raman medium first reduces $E_N$ due to the admixture of uncorrelated scattered photons. But when $n_{\mathrm{tot}} \simeq 1.5$, or $\Gamma_{\mathrm{cw}} \gtrsim 0.2$, this trend reverses and $E_N$ increases due to the generation of more quantum correlated photon pairs. 
At any $n_{\mathrm{tot}}$, the logarithmic negativity is larger than the one generated by spontaneous scattering, or by seeding either of the Stokes or the anti-Stokes sidebands, respectively. 
%It is also remarkable that seeding the anti-Stokes sideband appears more efficient to generate distillable entanglement than seeding on the Stokes side. 

We may also use seeding of the Raman sidebands to 'enhance' the nonclassicality, as observed in photon correlation measurements. This is shown in Fig.~\ref{fig:seeded-nonclassicality}. Surprisingly, here the roles of the seeds are reversed. Seeding with a Stokes photon enhances the nonclassicality witness $R$. 
%also t any total photon number. 
When seeding with anti-Stokes or with the superposition state~(\ref{eq:1_c}), $R$ is enhanced only close to $n_{\mathrm{tot}} = 1$, i.e. only when the Raman scattering is extremely weak. At larger photon numbers, the resulting $R$-value may even drop below the classical threshold. 

\subsubsection{Mean field dynamics}
%We finally turn to the nonreciprocal nature of Raman scattering. 

\begin{figure}
    \centering
    \includegraphics[trim=0cm 7cm 0cm 0cm, clip, width=1\linewidth]{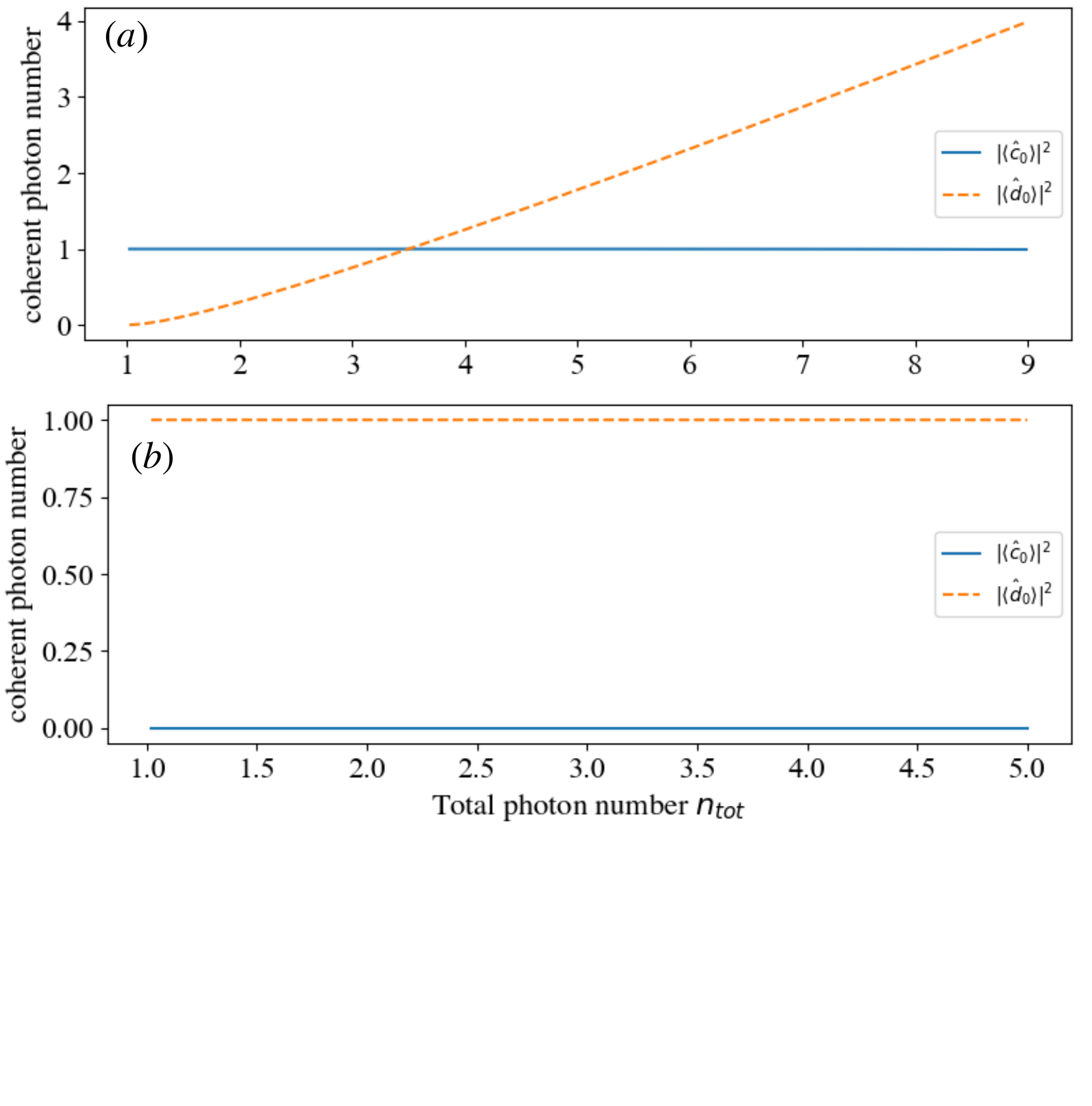}
    \caption{(a) Coherent state amplitudes of eigenmodes $\hat{c}_0$ and $\hat{d}_0$, i.e. $|\langle \hat{c}_0 \rangle|^2$ and $|\langle \hat{d}_0 \rangle|^2$, with initial state~(\ref{eq:c-coherent}). 
    (b) same as for (a) with initial state~(\ref{eq:d-coherent}).}
    \label{fig:dark-state}
\end{figure}

The nonreciprocal nature of Raman scattering can also be witnessed on a mean field level: 
Using Eqs.~(\ref{eq:auxilliary}) and (\ref{eq.c-Lindbladian}), and defining the dimensionless Bogoliubov-like operator $\hat{c}_0 = (\hat{a}_{\mathrm{aS}} + \hat{a}_{\mathrm{St}}^\dagger) / \sqrt{2}$, we find 
\begin{align}
    \partial_x \langle \hat{c}_0\rangle = 0.
\end{align}
In other words, its expectation value is a constant of motion in spontaneous Raman scattering. 
Note that $\hat{c}_0$ is not Hermitian, i.e. not an observable. 
We further define the conjugate ''momentum" $\hat{d}_0 = - i (\hat{a}_{\mathrm{aS}} - \hat{a}^\dagger_{\mathrm{St}}) / \sqrt{2}$, which satisfies $[\hat{c}_0, \hat{d}^\dagger_0] = i$. Its mean field evolution reads %analogously
\begin{align}
    \partial_x \langle \hat{d}_0 \rangle = g_{\mathrm{NR}} \langle \hat{c}_0 \rangle, \label{eq:d-evolution}
\end{align}
where $g_{\mathrm{NR}} = - 2 h_{\mathrm{St-aS}} (1 + 2 n_{\mathrm{th}}) + i \gamma_{\mathrm{St-aS}}$. 
Hence, the output coherent amplitude readily evaluates to $\langle \hat{d}_0 \rangle_{\mathrm{out}} = \langle \hat{d}_0 \rangle_{\mathrm{in}} + g_{\mathrm{NR}} \langle \hat{c}_0 \rangle_{\mathrm{in}}$. 
In other words, a coherent population in the $\hat{c}_0$-operator acts as a catalyst that can amplify the population of the $\hat{d}_0$-mode. Conversely, any initial population in the $\hat{d}_0$-mode does not stimulate coherent emission into the $\hat{c}_0$-mode. 
Remarkably, this is true not only at zero temperature, but at any mean phonon number. 
Since, $\hat{c}_0$ and $\hat{d}_0$ are not Hermitian, we need to specify how to initiate them. 
%it is somewhat tricky to initialize them: 
The amplitude of a coherent state prepared in the form 
\begin{align}
    \vert \psi_{\mathcal{C}} \rangle &= \exp ( \mathcal{C} \hat{d}_0^\dagger - \mathcal{C}^\ast \hat{d}_0) \vert 0 \rangle, \label{eq:c-coherent}
\end{align}
where $\mathcal{C} \in \mathbb{C}$, 
gives rise to $\langle \psi_{\mathcal{C}} \vert \hat{c}_0 \vert \psi_{\mathcal{C}} \rangle = \mathcal{C}$ and $\langle \psi_{\mathcal{C}} \vert \hat{d}_0 \vert \psi_{\mathcal{C}} \rangle = 0$. %, i.e. it is an eigenstate of $\hat{c}_0$. 
Decomposing Eq.~(\ref{eq:c-coherent}) into Stokes and anti-Stokes modes, it corresponds to a coherent state in the Stokes mode with amplitude $i \mathcal{C}^\ast / \sqrt{2}$ and with amplitude $- i \mathcal{C} / \sqrt{2}$ in the anti-Stokes mode. 
Fig.~\ref{fig:dark-state}(a) simulates the evolution of Raman scattering seeded by Eq.~(\ref{eq:c-coherent}) with $\langle \hat{c}_0 \rangle_{\mathrm{in}} = 1$, plotted vs. the total mean emitted photon number. 
It verifies numerically that the coherent amplitude of the initial state~(\ref{eq:c-coherent}) is not affected by Raman scattering. 
However, it does stimulate a coherent population in $\langle \hat{d}_0 \rangle$ according to Eq.~(\ref{eq:d-evolution}). 
%while generating population in mode $\hat{d}$.

Likewise, we can define a state with coherent population in $\hat{d}_0$. It reads
\begin{align}
    \vert \psi_{\mathcal{D}} \rangle &= \exp \left(\mathcal{D} \hat{c}^\dagger_0 - \mathcal{D}^\ast \hat{c}_0 \right) \vert 0 \rangle, \label{eq:d-coherent}
\end{align}
where $\mathcal{D} \in \mathbb{C}$. 
It gives rise to $\langle \psi_\mathcal{D} \vert \hat{c}_0 \vert \psi_\mathcal{D} \rangle = 0$ and $\langle \psi_\mathcal{D} \vert \hat{d}_0 \vert \psi_\mathcal{D} \rangle = \mathcal{D}$. 
Transforming Eq.~(\ref{eq:d-coherent}) into Stokes and anti-Stokes bases, we obtain coherent states in each mode, with complex amplitude $- \mathcal{D}^\ast / \sqrt{2}$ in the Stokes field, and $\mathcal{D} / \sqrt{2}$ in the anti-Stokes field. 
%On top of this coherent amplitude, there is of course spontaneous scattering, but it does not affect the coherent amplitude.
Raman scattering with Eq.~(\ref{eq:d-coherent}) as initial state is simulated in Fig.~\ref{fig:dark-state}(d), showing that coherent populations are not affected by its presence. 

\section{Conclusions}
\label{sec:conclusions}
%Our approach allows us to combine the photonic dynamics with an advanced microscopic treatment of the molecular degrees of freedom. The price we have to pay is that our approach cannot account for phonon heating and describe the depletion of the pump pulses.

We have derived a fully quantum mechanical theory of spontaneous Raman scattering, demonstrating that the photonic dynamics may be described by a cascaded master equation. This derivation neglects sample heating due to phonon excitations, and therefore is limited to weak driving where such backaction effects can be neglected. 
This unusual master equation has not been described in the literature to the best of our knowledge. It emerges naturally from the microscopics of the scattering process. At zero temperature, the emission of a Stokes photon may trigger emission of an anti-Stokes photon, but the opposite process is inhibited by the missing necessary energy. %, but not vice versa. 
Consequently, the Stokes subsystem 'drives' the anti-Stokes one, but there is no feedback. At finite temperature, this imbalance persists, but thermal phonon population enables the opposite scattering process, such that the imbalance becomes less pronounced. 

The master equation formulation of spontaneous Raman scattering enabled us to investigate photon statistics and quantum correlations beyond the individual photon limit, and to explore the impact of nonreciprocal couplings on the transmitted light fields. 
In a two-mode scenario, where we, e.g.,  spectrally filter two Stokes and anti-Stokes frequency bins, we find that, although the Raman photon-number correlations may resemble those associated with two-mode squeezed states, genuine quadrature squeezing is absent: the field quadrature variances always remain above the vacuum threshold, i.e. there is no quantum-mechanical squeezing, due to the unavoidable presence of uncorrelated photons. Furthermore, when the Raman sidebands are seeded by single photons, we quantify the generated entanglement (via logarithmic negativity) and find a strong dependence on whether the Stokes of anti-Stokes sidebands are seeded. Seeding with superposition states emerges as the best strategy for optimising the logarithmic negativity. Interestingly, this enhancement of the logarithmic negativity may lead to photon correlations that appear 'more classical' in their photon correlations than before. 

Beyond these fundamental insights, this framework provides a simple route to calculate quantum-optical observables and counting statistics, and consequently also to evaluate  sensitivity bounds via the quantum Fisher Information in the context of quantum sensing applications that will be explored in future work. 
The framework is constructed such that it connects naturally to the established semiclassical theory~\cite{schlawin2025theoryquantumenhancedstimulatedraman, sorelli2025ultimateresolutionlimitscoherent}. Hence, more complex, realistic molecular response functions featuring multiple vibrational lines or samples containing multiple species can be readily included in the framework.

\begin{acknowledgments}
    I would like to thank Maxime Jacquet, Hilton B. de Aguiar, Drew Voitiv, and Kenneth Burch for their helpful feedback on the manuscript. 
    I acknowledge the financial support of the Cluster of Excellence ``CUI: Advanced Imaging of Matter'' of the Deutsche Forschungsgemeinschaft (DFG) --- EXC 2056 --- project ID 390715994 and of the DFG research unit 'FOR5750: OPTIMAL' - project ID 531215165. 
    This research was supported in part by grant NSF PHY-2309135 to the Kavli Institute for Theoretical Physics (KITP).
\end{acknowledgments}

%\begin{figure*}
%    \centering
%    \includegraphics[trim = 0 600 220 0, clip, width=0.9\linewidth]{R-vs-nth.pdf}
%    \caption{ Nonclassicality witness $R$ (blue), Eq.~(\ref{eq:R-factor}), and the perturbation expectation~(\ref{eq:R-theory}) (organge) are shown (a) in the isolated pair regime, where $\langle \hat{n}_{\mathrm{St}} \rangle \ll 1$, (b) in a transition regime,  $\langle \hat{n}_{\mathrm{St}} \rangle \sim 1$, and (c) in a high-intensity regime  $\langle \hat{n}_{\mathrm{St}} \rangle > 1$.}
%    \label{fig:R-factor}
%\end{figure*}

\appendix
\section{Field operators}
\label{sec:field-operator}
We consider a quasi-onedimensional geometry, as sketched in Fig.~\ref{fig:sketch}(a), and write the field operators as \cite{Loudon}
\begin{align} 
    \hat{E} (z, t) &= \int \frac{d\omega}{2\pi} \sqrt{\frac{\hbar \omega}{2 \epsilon_0 c A}} e^{i (k (\omega)z - \omega t)} \hat{a} (\omega),
\end{align}
where $c$ is the speed of light (in the material), and $A$ the transverse beam width. 
The relevant frequencies are centered around the frequency of the classical pump $\omega_{\mathrm{pu}}$, which allows us to invoke the narrow bandwidth approximation: we approximate the field normalization by its value at the pump frequency, $E_{\mathrm{vac}} = (\hbar \omega_{\mathrm{pu}} / 2 \epsilon_0 c A)$, and write
\begin{align} \label{eq:E-operator_def}
    \hat{E} (z, t) &\simeq E_{\mathrm{vac}} \int \frac{d\omega}{2\pi} e^{i (k (\omega)z - \omega t)} \hat{a} (\omega) \notag \\
    &= E_{\mathrm{vac}} \hat{a} (z, t).
\end{align}

\section{Derivation of the master equation}
\label{app:derivation}
\begin{figure*}
    \centering
    \includegraphics[width=0.9\linewidth]{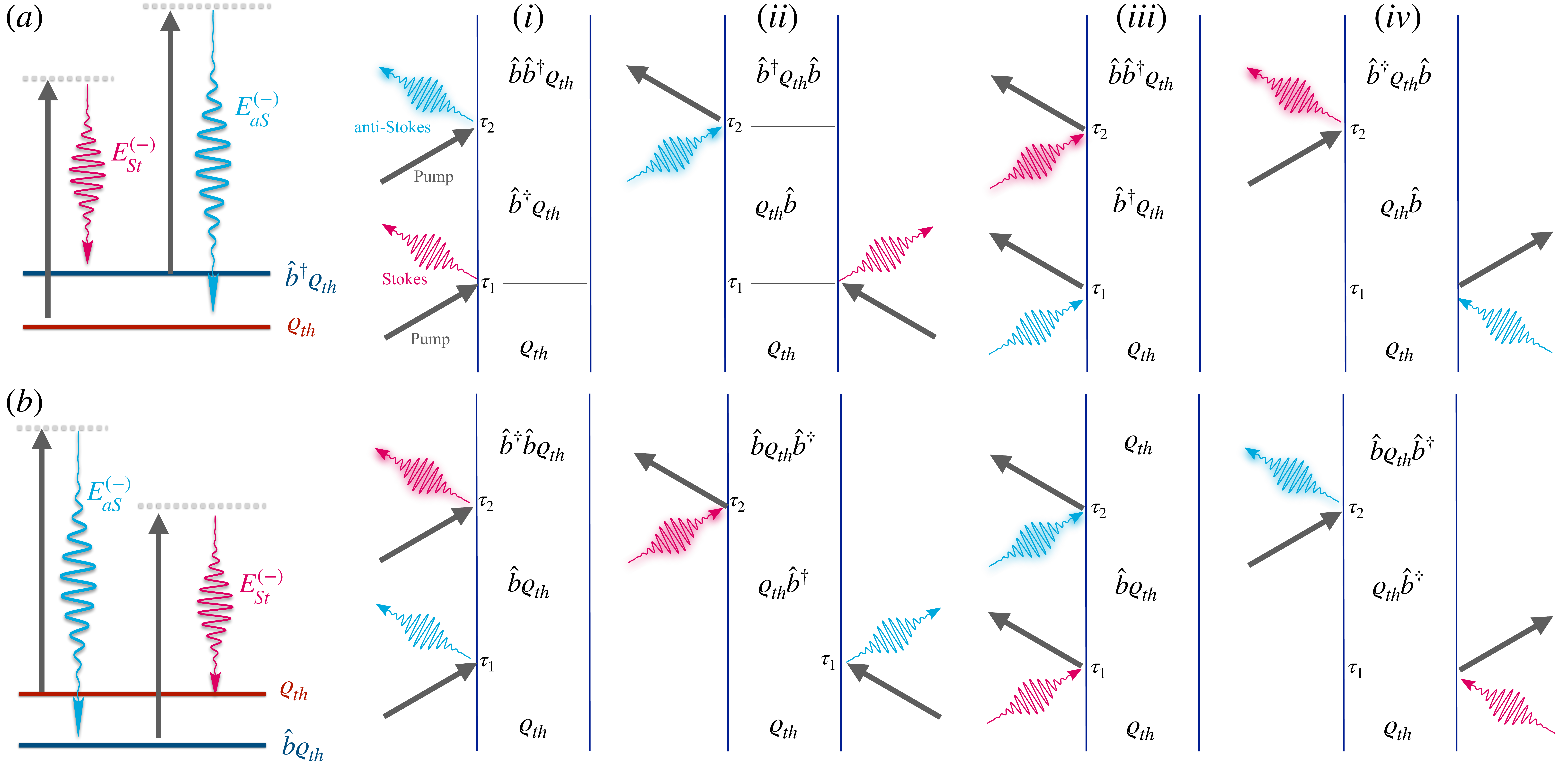}
    \caption{Feynman diagrams accounting for correlated Stokes anti-Stokes scattering. 
    (a) Diagrams describing phonon excitation, proportional to $n_{\mathrm{th}} + 1$. (b) Diagrams describing phonon annihilation, proportional to $n_{\mathrm{th}}$. }
    \label{fig:St-aS-diagrams}
\end{figure*}

\subsection{Feynman diagram rules}

The Feynman diagram rules are presented in~\cite{schlawin2025theoryquantumenhancedstimulatedraman, JCP2025_MDspec}, but for convenience we adapt them here.
\begin{itemize}
    \item The two vertical lines track the time evolution of the ket and bra side of the density matrix, respectively. 
    \item An red (blue) arrow pointing left with an accompanying solid line pointing in the opposite direction describes the action of the operator $ - i \hbar^{-1} \alpha_{\mathrm{R}} E_{\mathrm{pu}} \hat{E}_{\mathrm{St}} \hat{b}^\dagger$ ($- i \hbar^{-1} \alpha_{\mathrm{R}} E_{\mathrm{pu}} \hat{E}_{\mathrm{aS}} \hat{b}$), and vice versa. The operators are applied to the density matrix from the left (right), when they appear on the left (right) side of the diagram.
    \item Each operator acting on the bra side incurs a phase factor $-1$.  
    \item The molecule evolves freely in between these interaction events. 
    \item The final interaction is given by the superoperator $- i \hbar^{-1}\alpha_{\mathrm{R}} E_{\mathrm{pu}} \hat{E}_{St, -} \hat{b}^\dagger_{+}$ ($- i \hbar^{-1} \alpha_{\mathrm{R}} E_{\mathrm{pu}} \hat{E}_{aS, -} \hat{b}_+$), or its Hermitian conjugate.
    \item After the final interaction, the molecular degrees of freedom are traced out. Hence, diagrams which do not finish in a population state vanish. 
    \item The cumulant is obtained from time-ordered integrations of the light-matter interaction events.
\end{itemize}

\subsection{Correlated Stokes-anti-Stokes diagrams}

The molecules can be treated as point-like, such that the leading-order cumulants also remain local operators~\footnote{Higher-order cumulants may also include so-called cascading effects, where the emission of one molecule is reabsorbed by another. These will not be considered here. }. 
The diagrams involving both Raman sidebands are then given in Fig.~\ref{fig:St-aS-diagrams}. 
Using the Feynman diagram rules above, %in \cite{schlawin2025theoryquantumenhancedstimulatedraman}, 
these evaluate to superoperators acting on the photonic Hilbert space. 
For instance, diagram $(i)$ in Fig.~\ref{fig:St-aS-diagrams}(a) reads
\begin{widetext}
\begin{align} \label{eq:K_i}
    \mathcal{K}_{(i)} (\vec{r}) &= \left( - \frac{i}{\hbar} \right)^2 |\alpha_{\mathrm{R}}|^2 \int^{\infty}_{-\infty} \!\! d\tau_2 \int^{\tau_2}_{-\infty} \!\! d\tau_1 \; \langle \hat{b} (\tau_2) \hat{b}^\dagger (\tau_1) \rangle E_{\mathrm{pu}} (\vec{r}, \tau_2) E_{\mathrm{pu}} (\vec{r}, \tau_1) \notag \\
    &\times \hat{E}^\dagger_{\mathrm{aS}, -} (\vec{r}, \tau_2) \hat{E}^\dagger_{\mathrm{St}, L} (\vec{r}, \tau_1).
\end{align}
Here, the subscripts ''L" and ''R" denote action from the left (the right), i.e. $E_L X = E X$ and $E_{\mathrm{R}} X = X E$.
We write the phonon correlation function as (here we have added a phenomenological decay rate $\gamma$)
\begin{align}
    \langle \hat{b} (\tau_2) \hat{b}^\dagger (\tau_1) \rangle = (1 + n_{\mathrm{th}}) e^{- i (\omega_{\mathrm{ph}} - i \gamma) (\tau_2 - \tau_1)}.
\end{align}
Even though here we add the decay rate phenomenologically, the formalism allows us to straightforwardly include more sophisticated response functions for realistic molecular models at this point. 
This yields the superoperator
\begin{align}
    \mathcal{K}_{(i)} (\vec{r})&= - \frac{ i|\alpha_{\mathrm{R}}|^2 (n_{\mathrm{th}} + 1)}{\hbar^2} \int \frac{d\omega_-}{2\pi} \frac{1}{\omega_- - \omega_{\mathrm{ph}} + i \gamma} \epsilon^\dagger_{\mathrm{aS}, -} (\vec{r}, \omega_-) \epsilon^\dagger_{\mathrm{St}, L} (\vec{r}, \omega_-).
\end{align}
%where we defined
With a similar evaluation of the other Feynman diagrams in Fig.~\ref{fig:St-aS-diagrams}, we arrive at
\begin{align} \label{eq.K_St-aS_full}
    \mathcal{K}_{\mathrm{St-aS}} (\vec{r}) \hat{\varrho} &= \left(\mathcal{K}_{(i)} (\vec{r}) +\mathcal{K}_{(ii)} (\vec{r}) +\mathcal{K}_{(iii)} (\vec{r})+\mathcal{K}_{(iv)} (\vec{r}) \right) \hat{\varrho}\notag \\
    &= \int \frac{d\omega_-}{2\pi} \left( R_{\mathrm{St-aS}} (n_{\mathrm{th}}+1) \bigg[ \epsilon^\dagger_{\mathrm{aS}}, \epsilon^\dagger_{\mathrm{St}} \hat{\varrho} \right]  -   R^\ast_{\mathrm{St-aS}} (n_{\mathrm{th}}+1) \left[ \epsilon_{\mathrm{aS}}, \hat{\varrho} \epsilon_{\mathrm{St}} \right]\notag \\
    &\qquad + R_{\mathrm{St-aS}} (n_{\mathrm{th}} + 1) \left[ \epsilon_{\mathrm{St}}, \epsilon_{\mathrm{aS}} \hat{\varrho} \right] - R^\ast_{\mathrm{St-aS}} (n_{\mathrm{th}} + 1) \left[ \epsilon^\dagger_{\mathrm{St}}, \hat{\varrho} \epsilon_{\mathrm{aS}}^\dagger \right]\bigg) \notag \\
    &\qquad + (st \leftrightarrow aS \;\&\; n_{\mathrm{th}}+1 \rightarrow n_{\mathrm{th}}), 
\end{align}
\end{widetext}
where $R_{\mathrm{St-aS}}$ is given by Eq.~(\ref{eq:R_St-aS}) for $N = 1$, because we have not included ensemble effects yet.

The first two lines of Eq.~(\ref{eq.K_St-aS_full}) are derived from the Feynman diagrams $(i)-(iv)$ in Fig.~\ref{fig:K_St+K_aS}(a). They account for scattering events that transiently excite phonons in the molecule and scale as $n_{\mathrm{th}}+1$. 
The third line in Eq.~(\ref{eq.K_St-aS_full}) describes the Feynman diagrams in Fig.~\ref{fig:St-aS-diagrams}(b), which account for scattering events that transiently lower phonon population in the molecule. Hence, these processes are obtained from exchanging Stokes and anti-Stokes fields in the first two lines, and they scale with $n_{\mathrm{th}}$ (rather than $n_{\mathrm{th}}+1$).

Eq.~(\ref{fig:K_St+K_aS}) takes the form of a cascaded master equation. To obtain a Lindblad form, we have to account for individual gain and loss processes of each sideband. 

\subsection{Local dissipation}

\begin{figure*}
    \centering
    \includegraphics[width=0.9\linewidth]{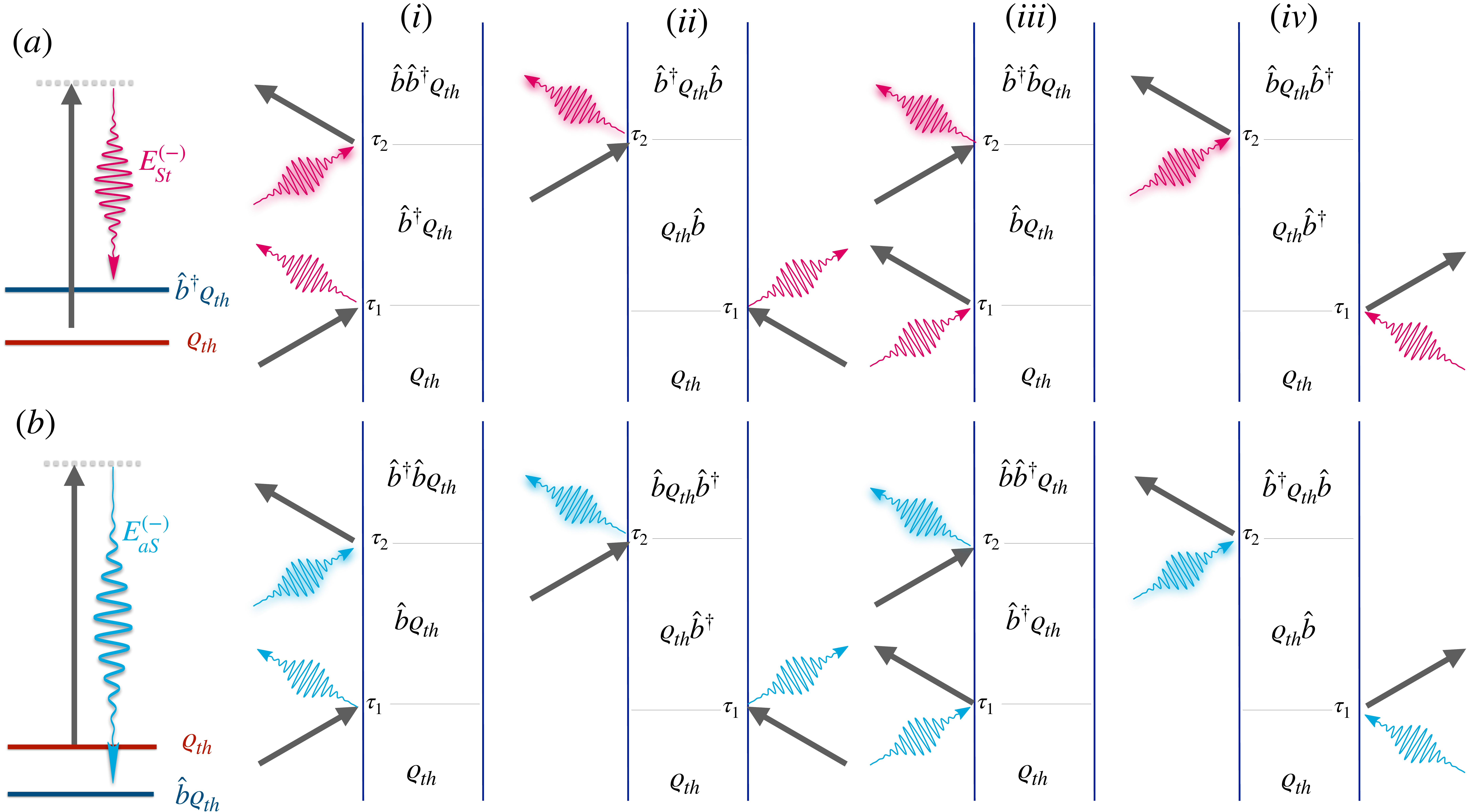}
    \caption{Feynman diagrams for Stokes and for anti-Stokes scattering. (a) Feynman diagrams for Stokes scattering. Diagrams $(i)$ and $(ii)$ describe Stokes gain, and scale as $n_{\mathrm{th}}+1$. Diagrams $(iii)$and $(iv)$ describe losses that reduce the phonon population ($\sim n_{\mathrm{th}}$).
    (b) Feynman diagrams for anti-Stokes scattering. Diagrams $(i)$ and $(ii)$ describe anti-Stokes gain, and scale as $n_{\mathrm{th}}$. Diagrams $(iii)$and $(iv)$ describe losses increasing the phonon population ($\sim n_{\mathrm{th}}+1$).
    }
    \label{fig:K_St+K_aS}
\end{figure*}

In addition to the correlated emission events described above, we also have to account for scattering events involving only one of the two siedbands. The corresponding Feynman diagrams are depicted in Fig.~\ref{fig:K_St+K_aS}, and translate into local Lindblad and Hamiltonian terms,
\begin{widetext}
\begin{align}
    \mathcal{K}_{\mathrm{St}} (\vec{r}) \hat{\varrho} &= \int \frac{d\omega_-}{2\pi} \left[ - i h_{\mathrm{St-aS}}(n_{\mathrm{th}}+1) \left[ \epsilon_{\mathrm{St}} \epsilon^\dagger_{\mathrm{St}}, \hat{\varrho} \right] + \gamma_{\mathrm{St-aS}} (n_{\mathrm{th}}+1) \mathcal{D}[\epsilon^\dagger_{\mathrm{St}}] \hat{\varrho} \right] \notag \\
    &+\int \frac{d\omega_-}{2\pi} \left[- i h_{\mathrm{St-aS}}n_{\mathrm{th}} \left[ \epsilon^\dagger_{\mathrm{St}} \epsilon_{\mathrm{St}}, \hat{\varrho} \right] + \gamma_{\mathrm{St-aS}} n_{\mathrm{th}} \mathcal{D}[\epsilon_{\mathrm{St}}] \hat{\varrho} \right], \label{eq.K_St}\\
    \mathcal{K}_{\mathrm{aS}} (\vec{r}) \hat{\varrho} &= \int \frac{d\omega_-}{2\pi} \left[- i h_{\mathrm{St-aS}} (n_{\mathrm{th}}+1)  \left[ \epsilon^\dagger_{\mathrm{aS}} \epsilon_{\mathrm{aS}}, \hat{\varrho} \right] + \gamma_{\mathrm{St-aS}} (n_{\mathrm{th}}+1) \mathcal{D}[\epsilon_{\mathrm{aS}}] \hat{\varrho} \right] \notag \\
    &+ \int \frac{d\omega_-}{2\pi} \left[ - i h_{\mathrm{St-aS}} n_{\mathrm{th}}  \left[ \epsilon_{\mathrm{aS}} \epsilon^\dagger_{\mathrm{aS}}, \hat{\varrho} \right] + \gamma_{\mathrm{St-aS}} n_{\mathrm{th}} \mathcal{D}[\epsilon^\dagger_{\mathrm{aS}}] \hat{\varrho} \right], \label{eq.K_aS}
\end{align}
\end{widetext}
where $\mathcal{D}[x] \hat{\varrho} = x \hat{\varrho} x^\dagger - \{ x^\dagger x, \hat{\varrho} \} / 2$ is a Lindblad form dissipator and $h_{\mathrm{St-aS}}$ and $\gamma_{\mathrm{St-aS}}$ are again given by Eq.~(\ref{eq:R_St-aS}) for $N = 1$
%\begin{align}
%    \mathcal{K}_{\mathrm{St}} (\vec{r}) &= \frac{ |\alpha_{\mathrm{R}}|^2 }{\hbar^2} \int \frac{d\omega_-}{2\pi} \left[ - i \frac{\Delta (n_{\mathrm{th}}+1)}{\Delta^2 + \gamma^2} \left[ \epsilon_{\mathrm{St}} \epsilon^\dagger_{\mathrm{St}}, \hat{\varrho} \right] + \frac{2\gamma (n_{\mathrm{th}}+1)}{\Delta^2 + \gamma^2} \left( \epsilon_{\mathrm{St}}^\dagger \hat{\varrho} \epsilon_{\mathrm{St}} - \frac{1}{2} \{ \epsilon_{\mathrm{St}} \epsilon^\dagger_{\mathrm{St}}, \hat{\varrho} \} \right) \right] \label{eq.K_gain}\\
%    \mathcal{K}_{\mathrm{aS}} (\vec{r}) &= \frac{ |\alpha_{\mathrm{R}}|^2 }{\hbar^2} \int \frac{d\omega_-}{2\pi} \left[ - i \frac{\Delta (n_{\mathrm{th}} +1)}{\Delta^2 + \gamma^2} \left[ \epsilon_{\mathrm{aS}}^\dagger \epsilon_{\mathrm{aS}}, \hat{\varrho} \right] + \frac{2\gamma (n_{\mathrm{th}}+1)}{\Delta^2 + \gamma^2} \left( \epsilon_{\mathrm{aS}} \hat{\varrho} \epsilon^\dagger_{\mathrm{aS}} - \frac{1}{2} \{ \epsilon^\dagger_{\mathrm{aS}} \epsilon_{\mathrm{aS}}, \hat{\varrho} \} \right) \right], \label{eq.K_loss}
%\end{align}
%with $\Delta =\omega_- - \omega_{\mathrm{ph}}$. 
Eq.~(\ref{eq.K_St}) describes the Stokes gain, scaling as $n_{\mathrm{th}}+1$, and Stokes loss, scaling as $n_{\mathrm{th}}$. 
Conversely, Eq.~(\ref{eq.K_aS}) accounts for anti-Stokes losses ($\sim n_{\mathrm{th}}+1$) and anti-Stokes gain  ($\sim n_{\mathrm{th}}$). 
\\

\subsection{Impulsive excitation}
\label{app:impulsive-excitation}
The dynamics induced by impulsive excitation are most conveniently evaluated in time domain. 
Inserting Eq.~(\ref{eq:impulsive-pump}) into the superoperator expression for diagram $(i)$ in Eq.~(\ref{eq:K_i}), we find 
\begin{align}
    \mathcal{K}_{(i)} &= - \gamma_{\mathrm{pulse}} (1 + n_{\mathrm{th}})\hat{a}^\dagger_{\mathrm{aS}, -} (t_0) \hat{a}_{\mathrm{St}, L} (t_0),
\end{align}
where $\gamma_{\mathrm{pulse}}$ is given in Eq.~(\ref{eq:gamma_pulse}). 
The same calculation can be readily repeated for all the Feynman diagrams. 

This result requires several comments: 
First, in the derivation, we obtain a phase factor $e^{- i 2\omega_{\mathrm{pu}} t_0}$ that can be absorbed into the definition of the time domain operators $\hat{a}^\dagger (t_0)$ to obtain slowly rotating operators as a function of $t_0$. Since the time $t_0$ has no physical relevance, we may set it to zero without loss of generality. 
Further, we note that $\gamma_{\mathrm{pulse}}$ has a different dimension compared to $\gamma_{\mathrm{St-aS}}$, due to the dimensionality of frequency and time domain operators, $\hat{a} (\omega)$ and $\hat{a} (t)$, in the continuous frequency limit considered here~\cite{Loudon}.
\\

\subsection{Propagation effects}
\label{sec:propagation-effects}
%$\mathcal{K}_{\mathrm{St}}$ and $\mathcal{K}_{\mathrm{aS}}$ always phase-matched 
%-> full dynamics with increased local losses, still of Lindblad form
We next consider the impact of propagation effects on the quantum map $\Phi$. In particular, we focus on collinear propagation as sketched in Fig.~\ref{fig:sketch}(a). 
We assume a constant molecular density $d_{\mathrm{mol}}$, and, as described at the beginning of this Appendix, point-like molecules, such that the cumulants are local. Hence, the full signal is generated by the cumulant
\begin{align}
    \mathcal{K}_{\mathrm{full}} &= d_{\mathrm{mol}}\int_{-L/2}^{L/2} \!\!\! dz \; \mathcal{K} (\vec{r}),
\end{align}
where $\mathcal{K} (\vec{r})$ is defined in Eq.~(\ref{eq.full-cumulant}). 
\\

\subsubsection{cw pump}
Using the two-photon operators~(\ref{eq:epsilon_St}) and (\ref{eq:epsilon_aS}), we write the correlated emission contribution at position $\vec{r} = (0,0,z)$ explicitly as
\begin{widetext}
\begin{align}
    \mathcal{K}_{\mathrm{St-aS}} (\vec{r}) \hat{\varrho}&=  E_{\mathrm{pu}}^2 E_{\mathrm{vac}}^2 (n_{\mathrm{th}}+1) \int \frac{d\omega_-}{2\pi} \notag \\
    &\qquad \bigg( R_{\mathrm{St-aS}}  e^{i \Delta k z} \left[ \hat{a}^\dagger_{\mathrm{aS}}, \hat{a}^\dagger_{\mathrm{St}} \hat{\varrho} \right] - R^\ast_{\mathrm{St-aS}}e^{- i \Delta k z} \left[ \hat{a}_{\mathrm{aS}}, \hat{\varrho} \hat{a}_{\mathrm{St}} \right]\notag \\
    &\qquad + R_{\mathrm{St-aS}}e^{i \Delta k z}  \left[ \hat{a}_{\mathrm{St}}, \hat{a}_{\mathrm{aS}} \hat{\varrho} \right] - R^\ast_{\mathrm{St-aS}} e^{- i \Delta k z} \left[ \hat{a}^\dagger_{\mathrm{St}}, \hat{\varrho} \hat{a}_{\mathrm{aS}}^\dagger \right]\bigg) \notag \\
    &\qquad + (st \leftrightarrow aS \;\&\; n_{\mathrm{th}}+1 \rightarrow n_{\mathrm{th}})
\end{align}
where $\hat{a}_{\mathrm{St}} = \hat{a}_{\mathrm{St}} (\omega_{\mathrm{pu}}-\omega_-)$ and $\hat{a}_{\mathrm{aS}} = \hat{a}_{\mathrm{aS}} (\omega_{\mathrm{pu}} +\omega_-)$. In addition, we have introduced the phase mismatch $\Delta k = 2 k_{\mathrm{pu}} (\omega_{\mathrm{pu}}) - k_{\mathrm{St}} (\omega_{\mathrm{pu}} - \omega_-) - k_{\mathrm{aS}} (\omega_{\mathrm{pu}} +\omega_-)$.
The integral over space simply yields 
\begin{align}
    \int_{-L/2}^{L/2} \!\!\! dz \; e^{i \Delta k z} &= L \text{sinc} \left( \frac{\Delta k L}{2} \right),
\end{align}
and we obtain the total correlated emission superoperator
\begin{align}
    \mathcal{K}_{\mathrm{St-aS}} \hat{\varrho}&=  N E_{\mathrm{pu}}^2 E_{\mathrm{vac}}^2 (n_{\mathrm{th}}+1) \int \frac{d\omega_-}{2\pi} \text{sinc} \left( \frac{\Delta k L}{2} \right)  \notag \\
    &\qquad \bigg( R_{\mathrm{St-aS}} \left[ \hat{a}^\dagger_{\mathrm{aS}}, \hat{a}^\dagger_{\mathrm{St}} \hat{\varrho} \right] - R^\ast_{\mathrm{St-aS}} \left[ \hat{a}_{\mathrm{aS}}, \hat{\varrho} \hat{a}_{\mathrm{St}} \right]\notag \\
    &\qquad + R_{\mathrm{St-aS}}  \left[ \hat{a}_{\mathrm{St}}, \hat{a}_{\mathrm{aS}} \hat{\varrho} \right] - R^\ast_{\mathrm{St-aS}} \left[ \hat{a}^\dagger_{\mathrm{St}}, \hat{\varrho} \hat{a}_{\mathrm{aS}}^\dagger \right]\bigg) \notag \\
    &\qquad + (st \leftrightarrow aS \;\&\; n_{\mathrm{th}}+1 \rightarrow n_{\mathrm{th}}),
\end{align}
\end{widetext}
where $N = d_{\mathrm{mol}} L$ is the total number of emitters.
Hence, the phase matching introduces the additional sinc-factor which reduces the scattering amplitude in frequency regimes with substantial phase mismatch $\Delta k L$. This mismatch depends on the group velocities inside the sample. In Eq.~(\ref{eq.full-cumulant}), this phase mismatch is included as $\eta (\omega_-) = \text{sinc} \left( \frac{\Delta k L}{2} \right)$. 

%The same is not true for the superoperators acting only on the Stokes or the anti-Stokes spaces, $\mathcal{K}_{\mathrm{St}}$ and $\mathcal{K}_{\mathrm{aS}}$. 
In contrast, the diagrams for $\mathcal{K}_{\mathrm{St}}$ and $\mathcal{K}_{\mathrm{aS}}$ do not suffer from phase mismatch, i.e. $\Delta k = 0$, and therefore we simply find
\begin{align}
    d_{\mathrm{mol}} \int_0^L \!\! dz \left( \mathcal{K}_{\mathrm{St}} (z) + \mathcal{K}_{\mathrm{aS}} (z) \right) &= N \left( \mathcal{K}_{\mathrm{St}} + \mathcal{K}_{\mathrm{aS}} \right).
\end{align}
Thus, these processes %are not restricted by phase mismatch, and 
always increase linearly with the number of molecules. 
Overall, this calculation shows that, since $\mathcal{K}_{\mathrm{St}}$ and $\mathcal{K}_{\mathrm{aS}}$ already have Lindblad form, the phase-matched dynamics are always described by a completely positive and trace preserving map.

\subsubsection{Broadband pump}

Here we go back to the definitions~(\ref{eq:epsilon_St}) and(\ref{eq:epsilon_aS}). Including their spatial dependence, we have, e.g.,
\begin{align}
    \hat{\epsilon}_{\mathrm{St}} (\omega_-) &= \int \frac{d\omega}{2\pi} E_{\mathrm{pu}} (\omega+\omega_-) \hat{E}_{\mathrm{St}} (\omega) e^{i q(\omega, \omega_-) z}, 
\end{align}
where $q (\omega, \omega_-) = - k_{\mathrm{pu}} (\omega+\omega_-) + k_{\mathrm{St}} (\omega)$. 
We likewise add the spatial dependence in the anti-Stokes field. 
With this definition, a similar calculation as in the cw case yields
\begin{widetext}
\begin{align}
    \mathcal{K}_{\mathrm{St-aS}}^{\mathrm{pulse}} \hat{\varrho} &= N E_{\mathrm{pu}}^2 E_{\mathrm{vac}}^2 (n_{\mathrm{th}}+1) \int \frac{d\omega_-}{2\pi} \int \frac{d\omega_1}{2\pi} \int \frac{d\omega_2}{2\pi} \notag \\
    &\times \text{sinc} \left( \frac{\Delta k(\omega_-;\omega_1, \omega_2) L}{2} \right) E_{\mathrm{pu}} (\omega_1 - \omega_-) E_{\mathrm{pu}} (\omega_2 +\omega_-) \notag \\
    &\bigg( R_{\mathrm{St-aS}} \left[ \hat{a}^\dagger_{\mathrm{aS}} (\omega_1), \hat{a}^\dagger_{\mathrm{St}} (\omega_2) \hat{\varrho} \right] - R^\ast_{\mathrm{St-aS}} \left[ \hat{a}_{\mathrm{aS}} (\omega_1), \hat{\varrho} \hat{a}_{\mathrm{St}} (\omega_2)\right]\notag \\
    &+ R_{\mathrm{St-aS}}  \left[ \hat{a}_{\mathrm{St}} (\omega_2), \hat{a}_{\mathrm{aS}} (\omega_1) \hat{\varrho} \right] - R^\ast_{\mathrm{St-aS}} \left[ \hat{a}^\dagger_{\mathrm{St}} (\omega_2), \hat{\varrho} \hat{a}_{\mathrm{aS}}^\dagger (\omega_1) \right]\bigg) \notag \\
    &\qquad + (st \leftrightarrow aS \;\&\; n_{\mathrm{th}}+1 \rightarrow n_{\mathrm{th}})
\end{align}
where the phase mismatch is now given by $\Delta k(\omega_-;\omega_1, \omega_2) = k_{\mathrm{pu}} (\omega_1 -\omega_-) + k_{\mathrm{pu}} (\omega_2 + \omega_-) - k_{\mathrm{aS}} (\omega_1) - k_{\mathrm{St}} (\omega_2)$. 

In the pulsed excitation scenario, $\mathcal{K}_{\mathrm{St}}$ and $\mathcal{K}_{\mathrm{aS}}$ are also affected. Here we only consider the dissipative contribution which is more important for the dynamics. The Hamiltonian terms may be treated analogously. 
We find explicitly
\begin{align}
    \mathcal{K}_{\mathrm{St}}^{\mathrm{pulse}} &= \int \frac{d\omega_-}{2\pi} \int \frac{d\omega_1}{2\pi} \int \frac{d\omega_2}{2\pi} E_{\mathrm{pu}} (\omega_1 + \omega_-) E_{\mathrm{pu}} (\omega_2 + \omega_-)  \text{sinc} \left(  \frac{\Delta k_{\mathrm{St}} (\omega_-; \omega_1, \omega_2) L}{2}\right) \notag \\
    &\times \bigg( \gamma_{\mathrm{St-aS}} (n_{\mathrm{th}}+1) \mathcal{D} [\hat{a}_{\mathrm{St}}^\dagger (\omega_1), \hat{a}_{\mathrm{St}} (\omega_2)] \hat{\varrho} + \gamma_{\mathrm{St-aS}} n_{\mathrm{th}} \mathcal{D} [\hat{a}_{\mathrm{St}} (\omega_2), \hat{a}^\dagger_{\mathrm{St}} (\omega_1)] \hat{\varrho}
    \bigg),
\end{align}
where $\Delta k_{\mathrm{St}} (\omega_-; \omega_1, \omega_2) = k_{\mathrm{pu}} (\omega_1 + \omega_-) - k_{\mathrm{pu}} (\omega_2 + \omega_-) - k_{\mathrm{St}} (\omega_1) + k_{\mathrm{St}} (\omega_2)$ and we defined the frequency-dependent 
\begin{align}
    \mathcal{D} [\hat{a}_{\mathrm{St}} (\omega_1), \hat{a}^\dagger_{\mathrm{St}} (\omega_2)] \hat{\varrho} = \hat{a}_{\mathrm{St}} (\omega_1) \hat{\varrho} \hat{a}_{\mathrm{St}}^\dagger (\omega_2) - \frac{1}{2} \left( \hat{a}_{\mathrm{St}}^\dagger (\omega_2) \hat{a}_{\mathrm{St}} (\omega_1) \hat{\varrho} + \hat{\varrho} \hat{a}_{\mathrm{St}}^\dagger (\omega_2) \hat{a}_{\mathrm{St}} (\omega_1) \right).
\end{align}
Likewise, the anti-Stokes superoperator evaluates to
\begin{align}
    \mathcal{K}_{\mathrm{aS}}^{\mathrm{pulse}} &= \int \frac{d\omega_-}{2\pi} \int \frac{d\omega_1}{2\pi} \int \frac{d\omega_2}{2\pi} E_{\mathrm{pu}} (\omega_1 - \omega_-) E_{\mathrm{pu}} (\omega_2 - \omega_-)  \text{sinc} \left(  \frac{\Delta k_{\mathrm{aS}} (\omega_-; \omega_1, \omega_2) L}{2}\right) \notag \\
    &\times \bigg( \gamma_{\mathrm{St-aS}} (n_{\mathrm{th}}+1) \mathcal{D} [\hat{a}_{\mathrm{aS}} (\omega_1), \hat{a}^\dagger_{\mathrm{aS}} (\omega_2)] \hat{\varrho} + \gamma_{\mathrm{St-aS}} n_{\mathrm{th}} \mathcal{D} [\hat{a}^\dagger_{\mathrm{aS}} (\omega_2), \hat{a}_{\mathrm{aS}} (\omega_1)] \hat{\varrho}
    \bigg),
\end{align}
with $\Delta k_{\mathrm{aS}} (\omega_-; \omega_1, \omega_2) = k_{\mathrm{pu}} (\omega_1-\omega_-) - k_{\mathrm{pu}} (\omega_2 - \omega_-) - k_{\mathrm{aS}} (\omega_1) + k_{\mathrm{aS}} (\omega_2)$. 
%These calculations show that 
\end{widetext}

%\begin{widetext}
%\begin{align}
%    \mathcal{K}_{\mathrm{St-aS}} (\vec{r}) &=  \gamma_{\mathrm{pulse}} (n_{\mathrm{th}}+1)\bigg( e^{i \Delta k z} \left[ \hat{a}^\dagger_{\mathrm{aS}} (t_0), \hat{a}^\dagger_{\mathrm{St}} \hat{\varrho} \right] - e^{- i \Delta k z} \left[ \hat{a}_{\mathrm{aS}}, \hat{\varrho} \hat{a}_{\mathrm{St}} \right]\notag \\
%    &\qquad\qquad\qquad\qquad  + e^{i \Delta k z}  \left[ \epsilon_{\mathrm{St}}, \epsilon_{\mathrm{aS}} \hat{\varrho} \right] - e^{- i \Delta k z} \left[ \hat{a}^\dagger_{\mathrm{St}}, \hat{\varrho} \hat{a}_{\mathrm{aS}}^\dagger \right]\bigg) \notag \\
%    &\qquad\qquad + (st \leftrightarrow aS \;\&\; n_{\mathrm{th}}+1 \rightarrow n_{\mathrm{th}})
%\end{align}
%\end{widetext}

\bibliography{sample}

\end{document}